\documentclass[onecolumn]{revtex4}

\usepackage{graphicx,rotating}
\usepackage{dcolumn}
\usepackage{bm}
\usepackage{epsfig}
\usepackage{longtable}
\usepackage{color}

\begin{document}

\title{Effect of graphene substrate on the spectroscopic properties of
photovoltaic molecules: role of the in-plane and out-of-plane $\pi$-bonds}

\author{Ma\l{}gorzata Wawrzyniak-Adamczewska}
\affiliation{%
Faculty of Physics, A. Mickiewicz University, ul. Umultowska 85, 61-614
Pozna\'n, Poland
}%

\author{Ma\l{}gorzata~Wierzbowska}\email{wierzbowska@ifpan.edu.pl}
\affiliation{%
Institut of Physics, Polish Academy of Sciences (PAS),
Al. Lotnik\'ow 32/46, 02-668 Warszawa, Poland
}%

\author{Juan Jos\'e Mel\'endez}\email{melendez@unex.es}
\affiliation{%
Department of Physics and Institute for Advanced Scientific Computing of Extremadura
(ICCAEX), University of Extremadura, Avenida de Elvas, s/n, 06006, Badajoz, Spain
}%

\begin{abstract}
The electronic structure of pentacene decorated with dipole groups (d-pentacene)
and adsorbed onto 
a graphene substrate has been studied within the density functional theory. 
Three reference configurations have been considered, 
namely the ideal molecule without distortions, 
the actual molecule including intramolecular distortions and 
the molecule adsorbed onto graphene. 
Calculations show a noticeable charge redistribution within the d-pentacene + graphene system 
due to molecular distortion, as well as the formation of weak $\pi$-bonds 
between the molecule and the substrate.
Additionally, the effect of the chemical modification of the terminal saturation 
with --H by --OH and =O is checked to explore the possibility of ``levels engineering''. 
The imaginary part of the dielectric function of d-pentacene in the ideal 
and distorted conformations and the molecule 
adsorbed at graphene were 
calculated within the random phase approximation. Results show that, even though molecular 
distortions change apreciably the absorption spectrum of isolated d-pentacene, 
the adsorbed molecule exhibits an optical spectrum which mimics quite much 
that of single graphene.
\end{abstract}


\maketitle

\section{Introduction}

Graphene is one of the frontier materials for flexible electronics, including photovoltaics. 
It has been used inside composite solar cells, either as the electrode  when
doped with N or B \cite{x1}  or as the optically active part \cite{x2}, 
as well as the hole transporting layer and even barrier for H$_2$O  in order to improve the 
resistivity against moisture \cite{x3}. 
As a substrate for the organic layers, graphene brings molecules into order \cite{x4},
and therefore sets rules for the molecular pattern \cite{x5}. 

When molecules with the dipole groups are adsorbed at graphene, 
they induce the electric field which shifts the electronic density of the graphene layer
according to the Stark effect. In the result, the dipole moment of about 2.1 Debye 
is induced in the graphene layer and its work function changes by about 1.5 eV \cite{x6} 
due to Wigner and Bardeen mechanism \cite{WB}.
The dipole groups of molecules tend to align in the ferroelectric order in 1D $\pi$-stacks
or 2D layers. Moreover, such ferroelectric organic layers possess other properties 
which are desired for applications in the solar devices.
One of them is a cascade alignment of the energy levels of subsequent molecular layers, 
both in the valence-band and conduction-band manifold \cite{x6}. 
These cascade states simultaneously play a role
of the donors or acceptors, depending on a direction from which the carriers arrive.
Importantly, the $\pi$-stacks of these molecules transport electrons and holes through
different parts in space, the mesogenic aromatic rings and the terminal dipole groups, 
respectively \cite{x7}. Thus, such separation of the carrier paths allows to decrease 
the charge recombination, which is the main problem restricting the power conversion efficiency.
   
In this theoretical work, we study the effect of a graphene substrate 
on the absorption spectra of pentacene decorated with the --CH$_2$CN and --COOH dipole groups. 
The optical properties of undoped pentacene were studied with 
\emph{ab initio} methods by other
authors in relation to the charge transfer between the molecule and a TiO$_2$ 
substrate \cite{x8}. It has been reported that this molecule has a HOMO--LUMO gap within 
the range of solar radiation and that the dipole groups do not change much this gap \cite{x9}. 
On the other hand, experimental and theoretical studies about 2D organic networks 
at highly oriented pyrolytic graphite (HOPG) surface, which is very similar to graphene,
reported the possibility to achieve tunable band gaps by a modification of the number 
of the aromatic rings \cite{x10}.    
For these reasons, the chosen molecule is a very good candidate to build organic layers for 
efficient photovoltaic cells with a graphene transparent electrode.

\section{Materials and Methods}

We considered the pentacene molecule decorated with four COOH and six CH$_2$CN dipole groups, 
with the rest of bonds saturated with hydrogen; in what follows, we will refer to this molecule 
as d-pentacene. It will be shown below that the optical spectrum of the molecule depends on 
both its own geometry and its electrostatic interactions with the graphene layer. 
To study these effects, we considered three arrangements: 
1) plane molecule, with an ``ideal" geometry optimized in vacuum (Pi),
2) the molecule calculated in vacuum but with the "relaxed" geometry, 
 assumed to be the same as that of the molecule placed on top of the graphene layer (Pr), 
and 3) the molecule with the geometry optimized on top of the graphene layer (PrG). 

Density functional theory (DFT) calculations have been performed using the plane-wave 
pseudopotential code Quantum ESPRESSO \cite{QE}. The Perdew-Burke-Ernzerhof parametrization for 
the exchange-correlation functional has been used \cite{PBE}. Norm-conserving pseudopotentials 
with a plane-wave energy cutoff of 60 Ry have been used for all atomic species. 
The uniform $2\times 2\times 1$ Monkhorst-Pack k-mesh \cite{MP} sufficed to sample the first Brillouin 
zone for the almost-square supercells, of planar side of around 20 $\AA$; a comparison between
this mesh and $6\times 6\times 1$ k-mesh for the projected DOS is reported in Figure S1 in the  
supporting information.
A vacuum of 30 $\AA$ separated the periodic slabs. The geometry optimization was achieved by 
allowing the graphene + molecule system to relax following the Broyden-Fletcher-Goldfarb-Shanno 
algorithm \cite{BFGS} until the forces on each atom decreased below $1\times 10^{-3}$ eV/$\AA$.

The dielectric matrix has been calculated, in the linear response regime and within 
the random phase approximation (RPA), using the Yambo code \cite{yambo}. In all cases, local field effects 
up to 14 eV were considered for the response function. Calculations of the dielectric matrix 
involved 1000 valence + conduction bands. 
A damping energy parameter of 0.1 eV was used to mimic experimental peak widths. 
Polarization of light was assumed to be polarized along the [110] direction, parallel to the graphene layer.  

\section{Results and Discussion}

\subsection{Geometric Characteristics}

\begin{figure}
\centerline{
       \includegraphics[scale=0.26,angle=0.0]{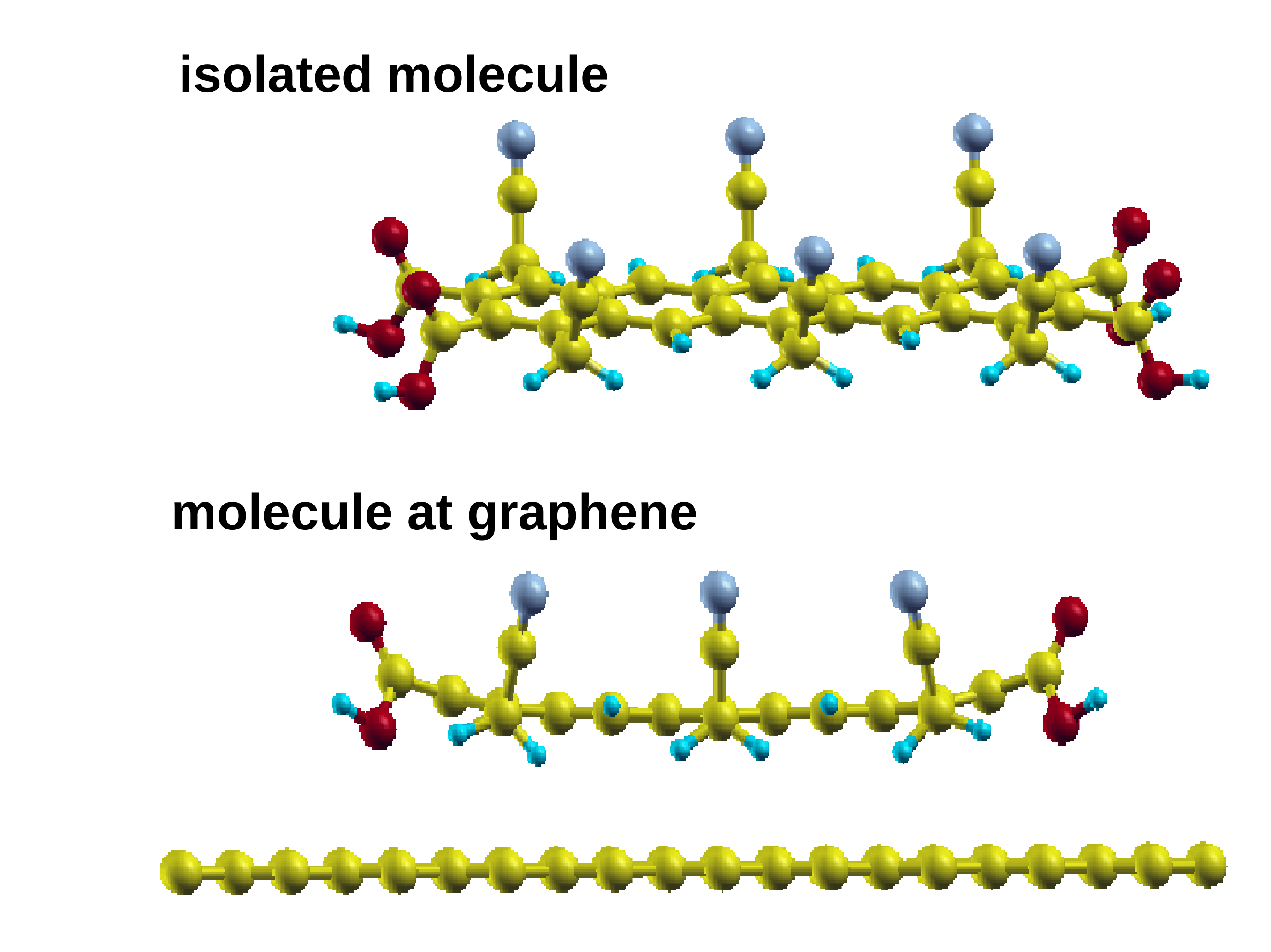}\label{Fig1-a}
       \includegraphics[scale=0.26,angle=0.0]{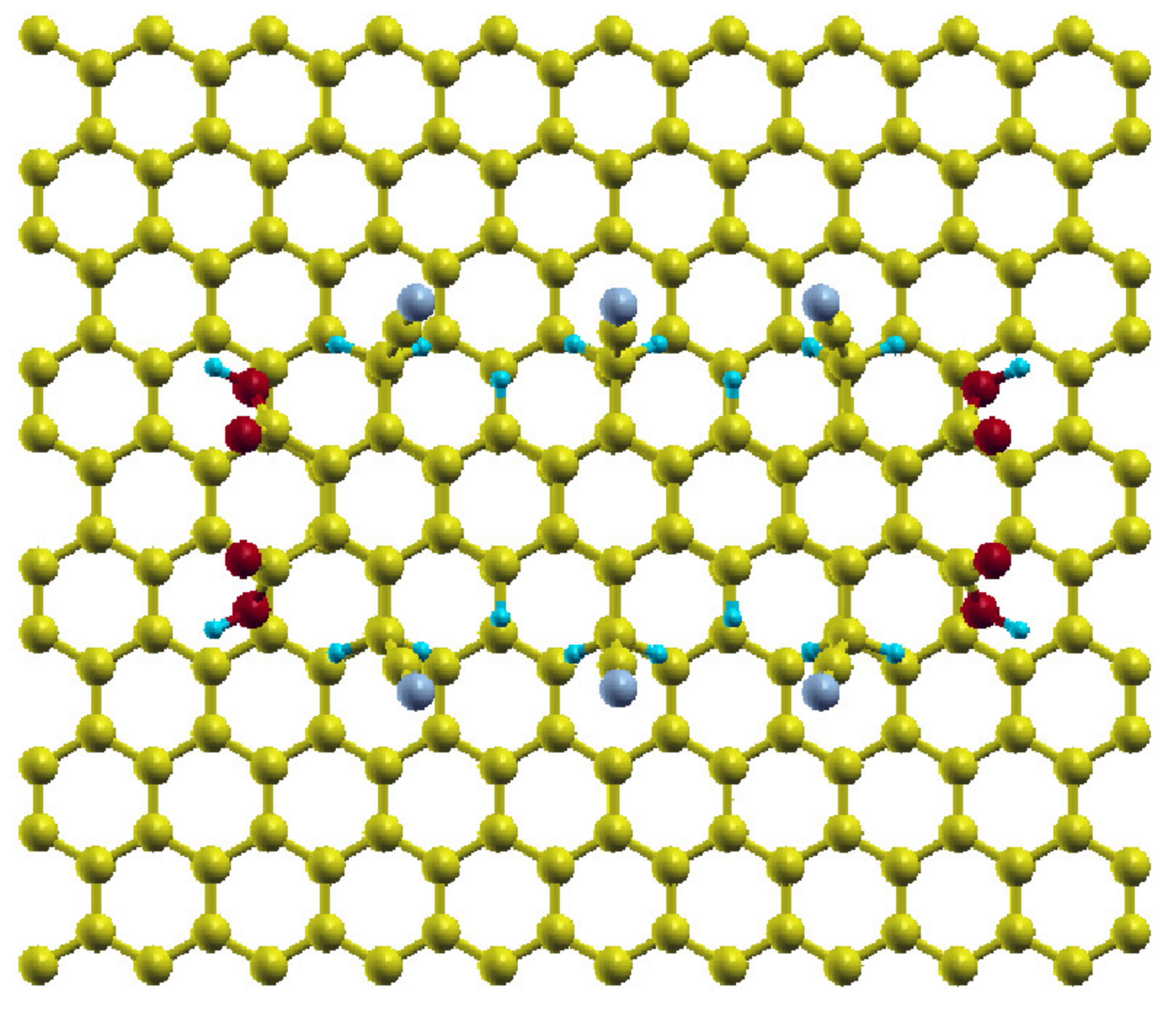}\label{Fig1-b}}
\caption{a) The atomic structure of an isolated d-pentacene molecule (top) and the molecule adsorbed on graphene (bottom). b) Top view of the latter. Carbon atoms appear in yellow, nitrogen and oxygen ones appear in grey and red, respectively, and hydrogen in cyan.}
        \label{Fig1}
\end{figure}

Figure \ref{Fig1-a} shows views of the isolated d-pentacene molecule (top) and that adsorbed at 
graphene (bottom); Figure \ref{Fig1-b} shows a top view of the adsorbed molecule. We thought primarily about the nature 
of the molecule adsorption, that is, whether it was chemisorption or any kind of physisorption. 
In principle, one would expect chemisorption to take place, provided that the molecule is 
close enough to the graphene layer. Thus, we placed initially a plane d-pentacene molecule close 
to graphene, with the C(ring)--C(graphene) and O(molecule)--C(graphene) distances being 
2.0 $\AA$ and 1.48 $\AA$, respectively. Besides, we located the hydrogen atoms of the --COOH 
groups far away from the graphene layer, to ease the O(pentacene)--C(graphene) bonding in case 
that it happened. After relaxation, we found the molecule to be repeled and distorted by graphene, 
with the distances between the central and outermost C atoms of d-pentacene and the graphene layer 
being 2.98 $\AA$ and 3.76 $\AA$, respectively. 

\begin{figure}[h!]
\centerline{
    \includegraphics[scale=0.125,angle=0.0]{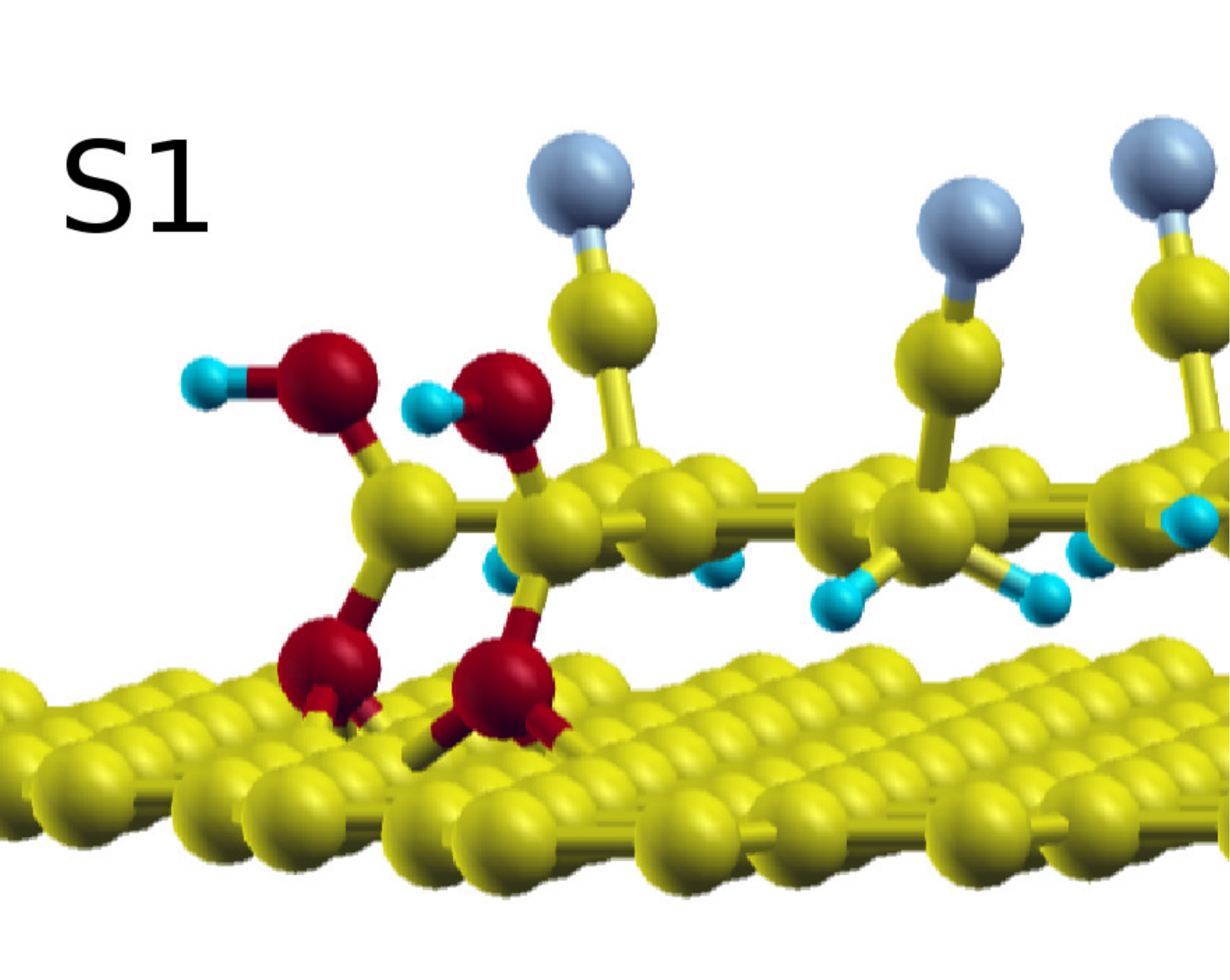} \hspace{0.5cm}
    \includegraphics[scale=0.125,angle=0.0]{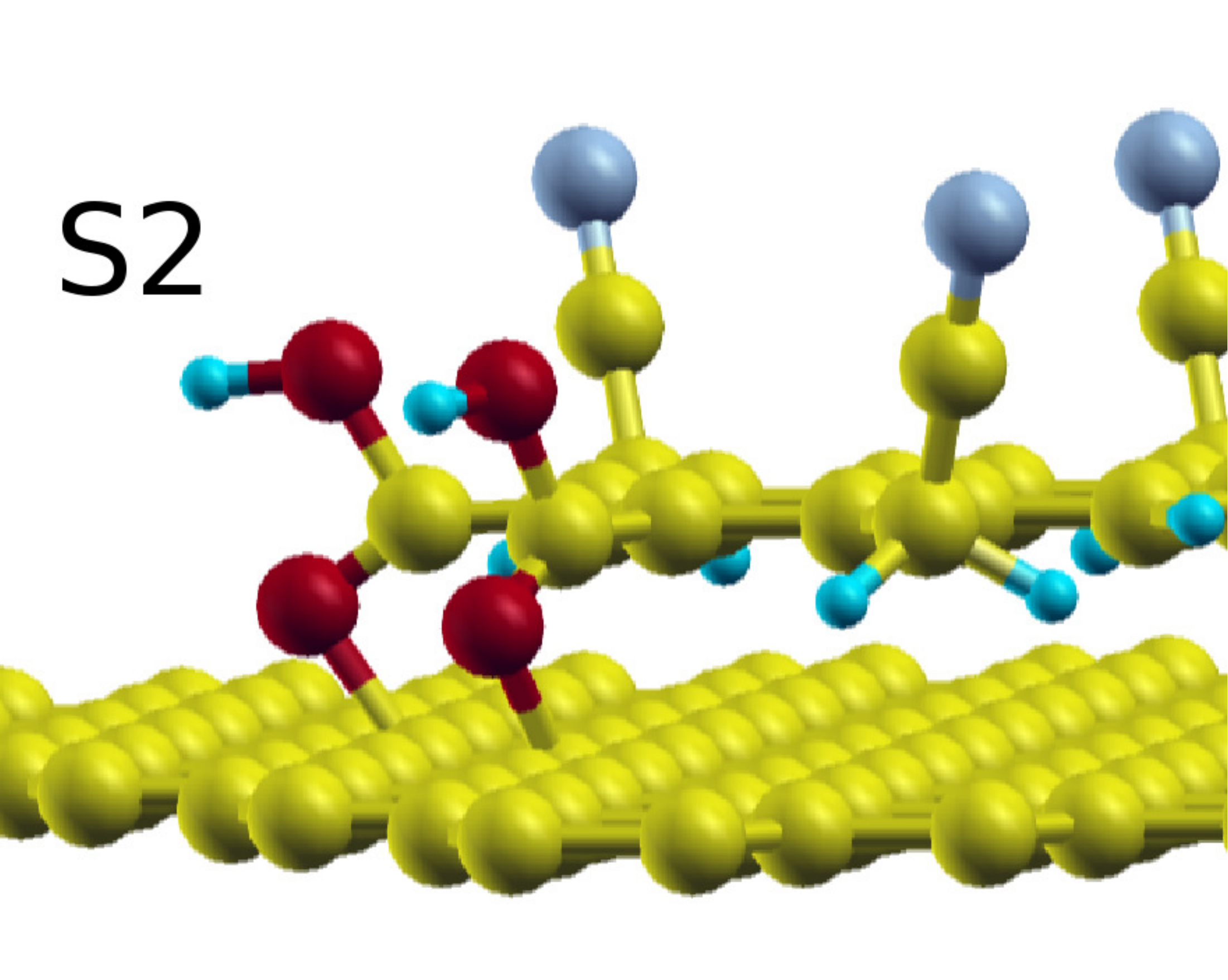} \hspace{0.5cm}
    \includegraphics[scale=0.125,angle=0.0]{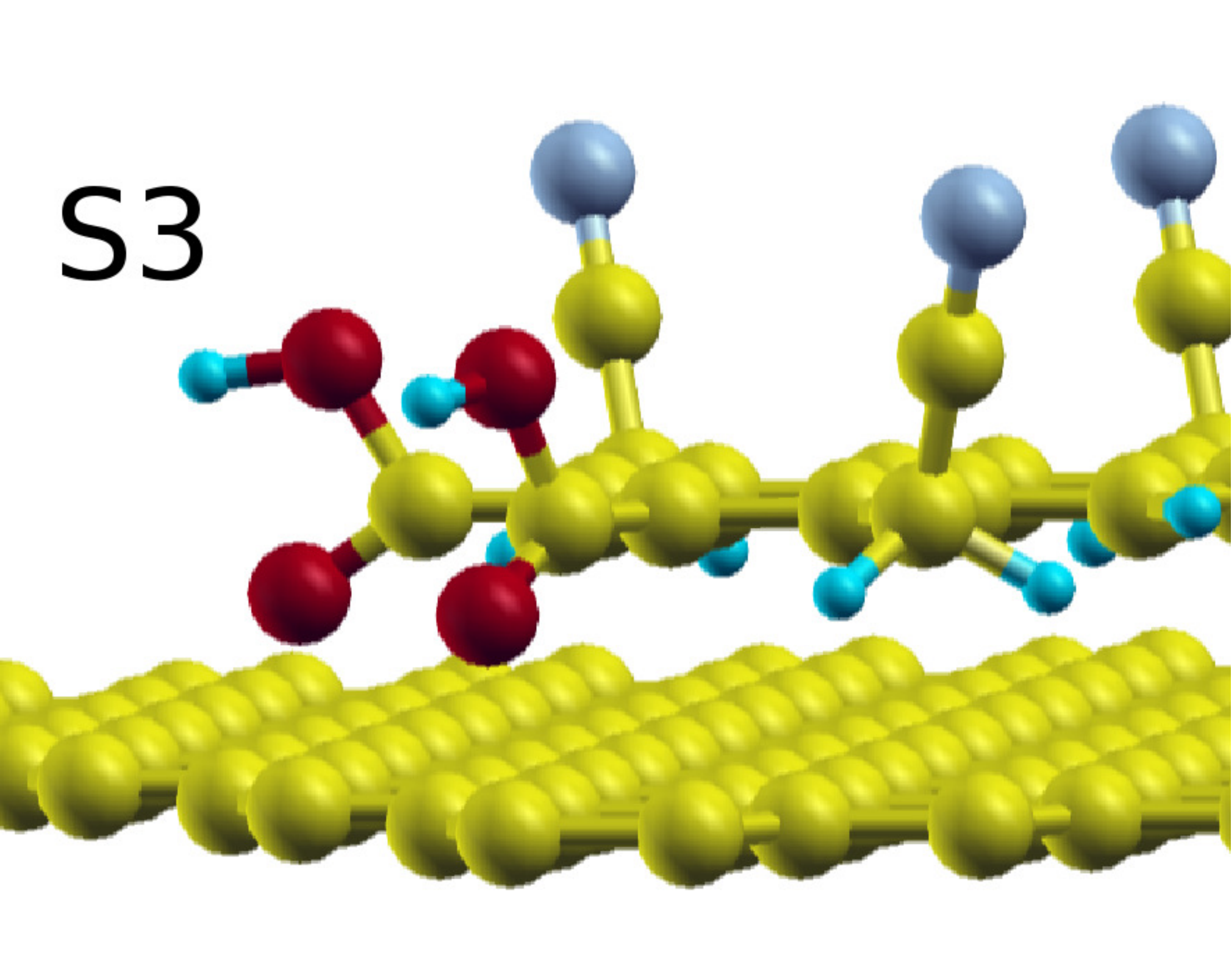}}
\centerline{
    \includegraphics[scale=0.125,angle=0.0]{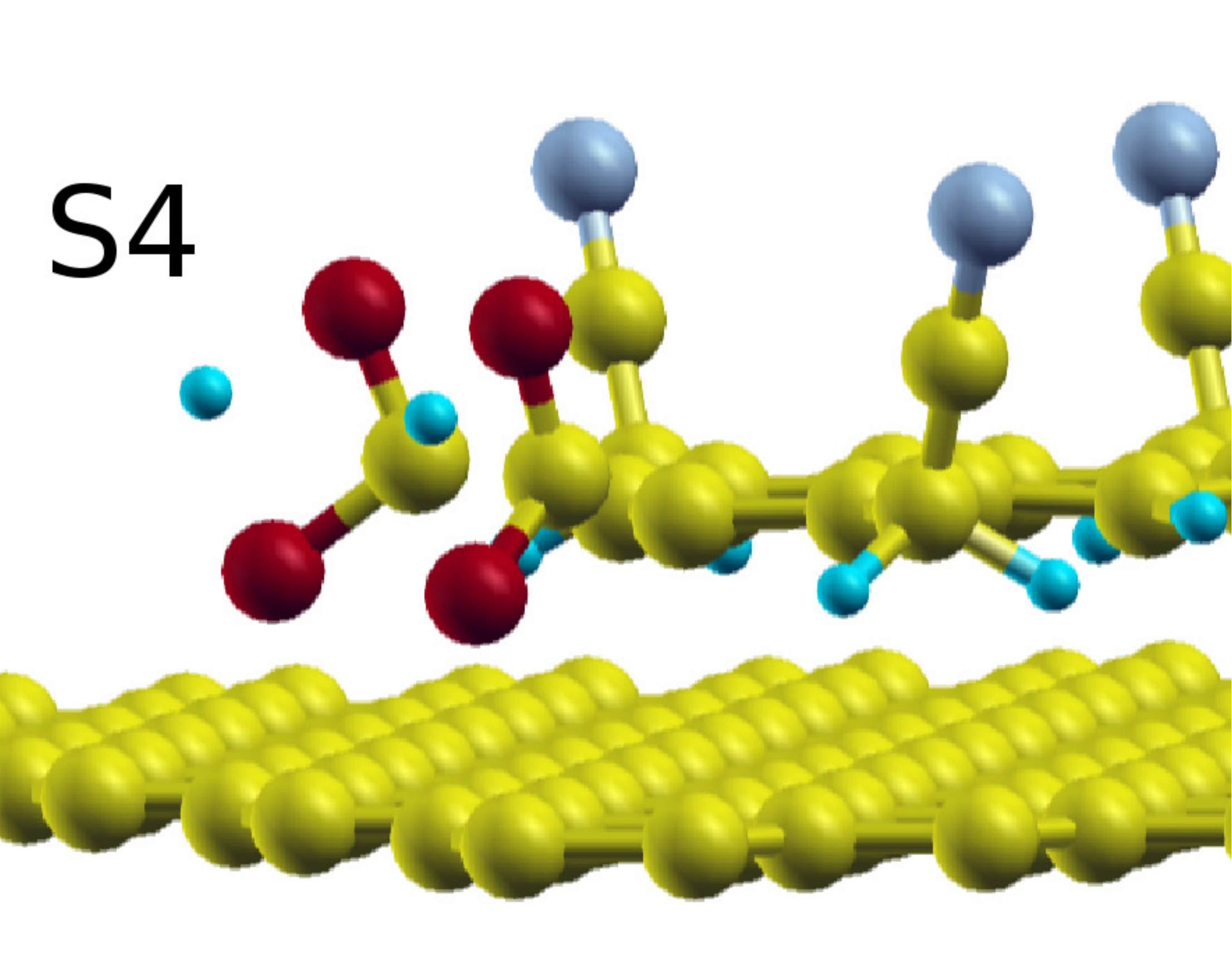} \hspace{0.5cm}
    \includegraphics[scale=0.125,angle=0.0]{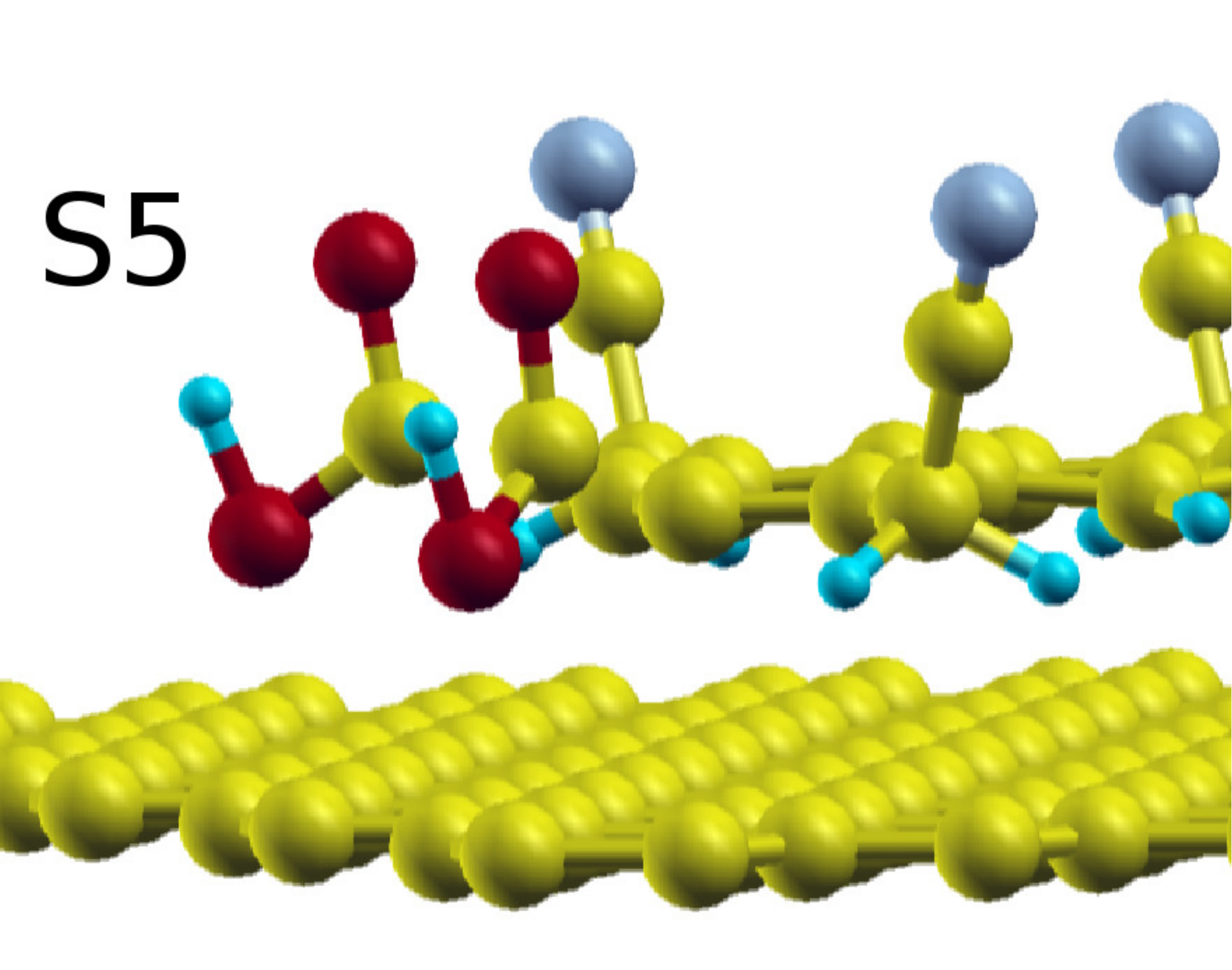} \hspace{0.5cm}
    \includegraphics[scale=0.125,angle=0.0]{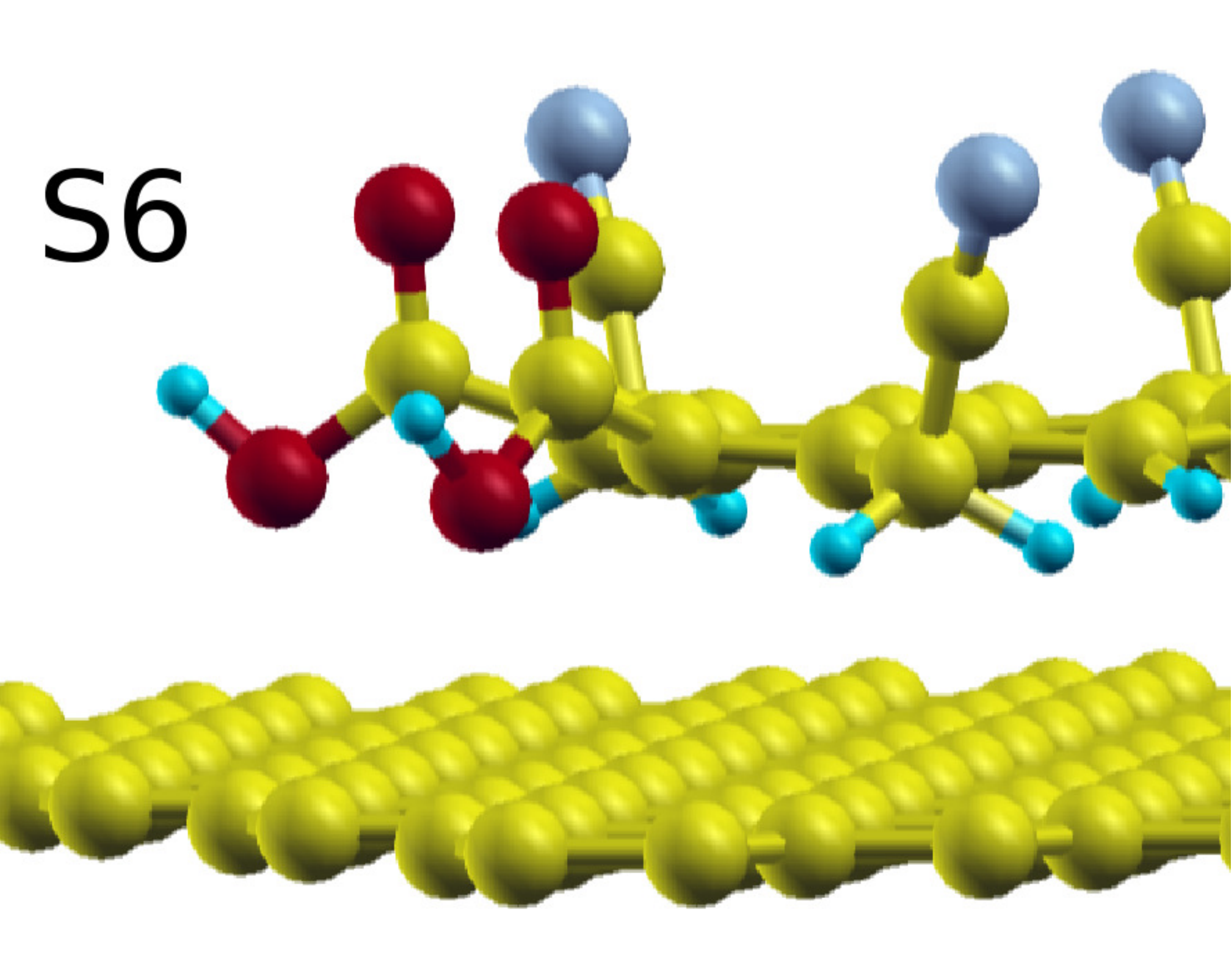}}

        \caption{The initial (S1) and five transient atomic structures
(S2--S6) during the optimization of the geometry of the molecule
adsorbed at graphene. The color code is the same as for Figure 1.}
        \label{Fig2}
\end{figure}

In addition, the hydrogen atoms from the --COOH groups swapped their positions, and got bound 
to the oxygen atoms closer to graphene. This way, the dipole moment of the --COOH aligned with 
that of the CH$_2$CN group, in agreement with a previous work where the ferroelectric 
structure of a similar molecule, with just one C-ring and the same dipoles, was lower in energy 
than any other configuration of dipoles \cite{x6}. 
The optimization process is schematized in Figure \ref{Fig2}, 
which displays details of the initial --COOH configuration as well as five transient stages of 
the atomic structure. It is interesting to inquiry about why hydrogen atoms swap their positions. 
In our opinion, this could be caused by the need of the system to align the different dipoles 
of the molecule. Indeed, the --CH$_2$CN groups have dipole moments, which induce an electric field 
in the molecule; the hydrogen atoms then swap to screen such a field. Incidentally, the fact 
that graphene is slightly negative would help it to attract hydrogens as well.
We checked that the geometry optimized with the method containing the self-interaction correction,
namely the pSIC approach \cite{y1,y2}, also swaps the hydrogens to place them at lower oxygens
of the COOH group.  

In conclusion, our results indicate that d-pentacene is physisorped on graphene, and that adsorption 
is accompanied by a distortion of the molecule, which adopt a ``cone"-like configuration with 
the --COOH groups further from graphene than the rest of the molecule.  

\subsection{Electronic Structure of Decorated Pentacene}

\begin{figure}[h!]
\centerline{
      \includegraphics[scale=0.25,angle=0.0]{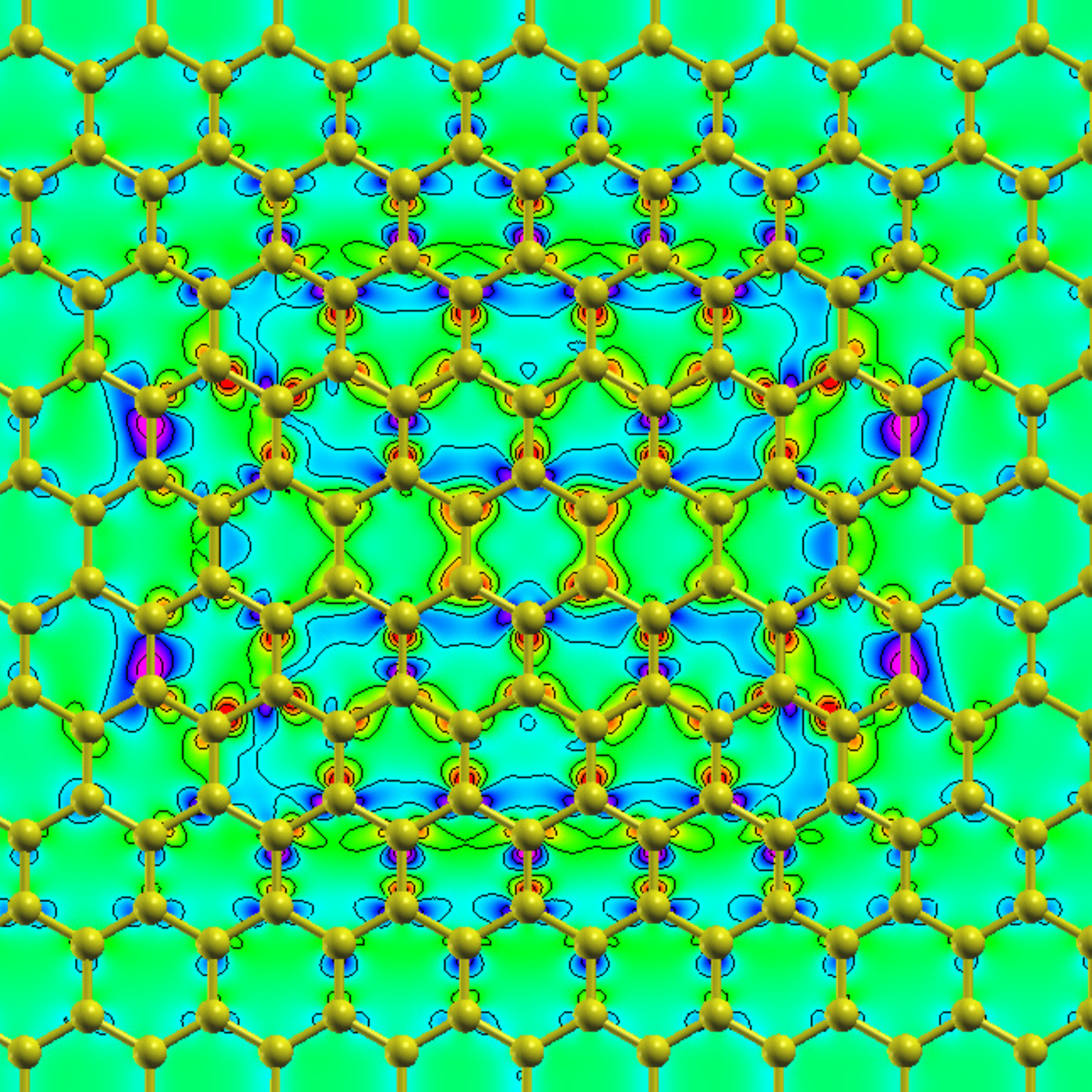}\label{F3a}
      \includegraphics[scale=0.25,angle=0.0]{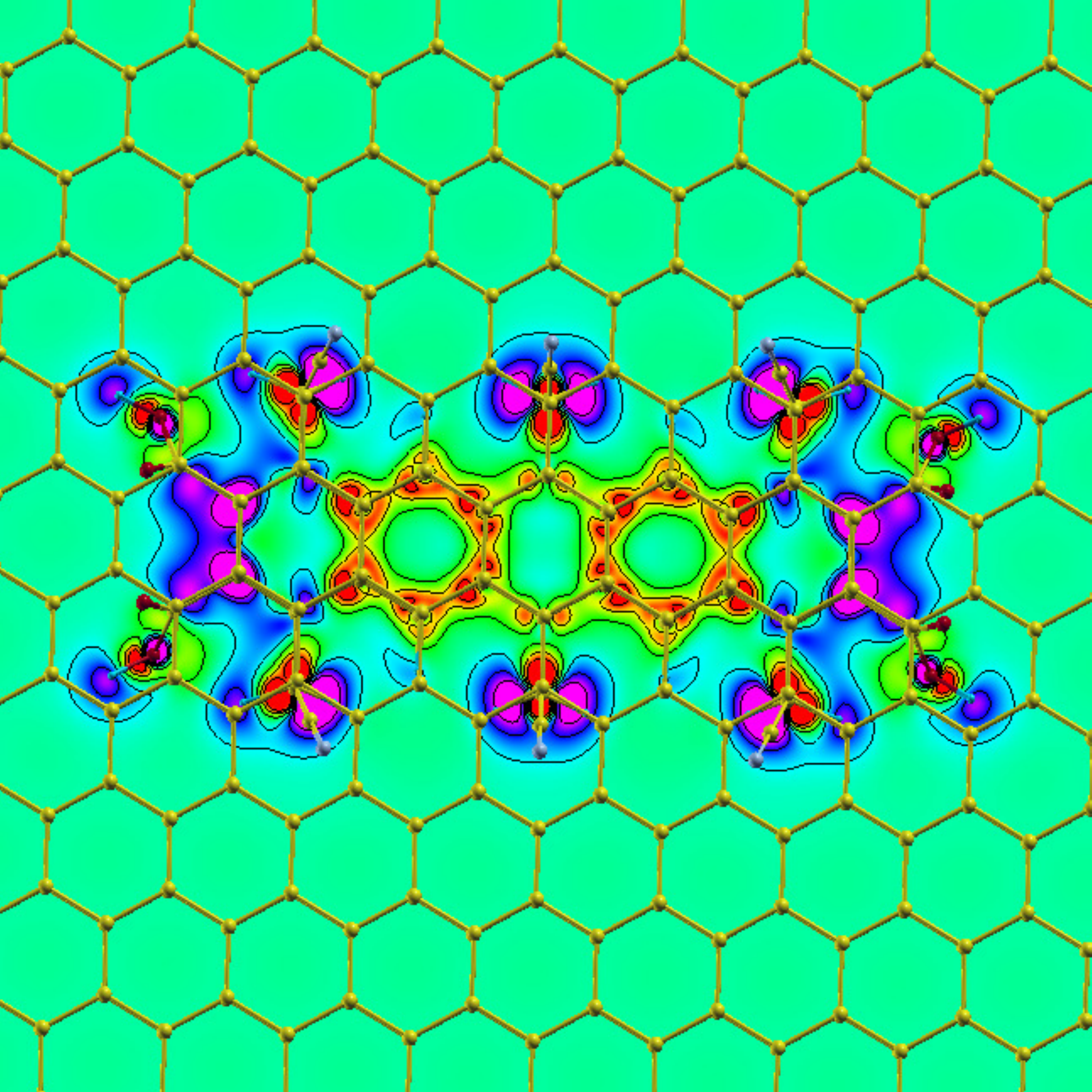}\label{F3b}
      \includegraphics[scale=0.26,angle=0.0]{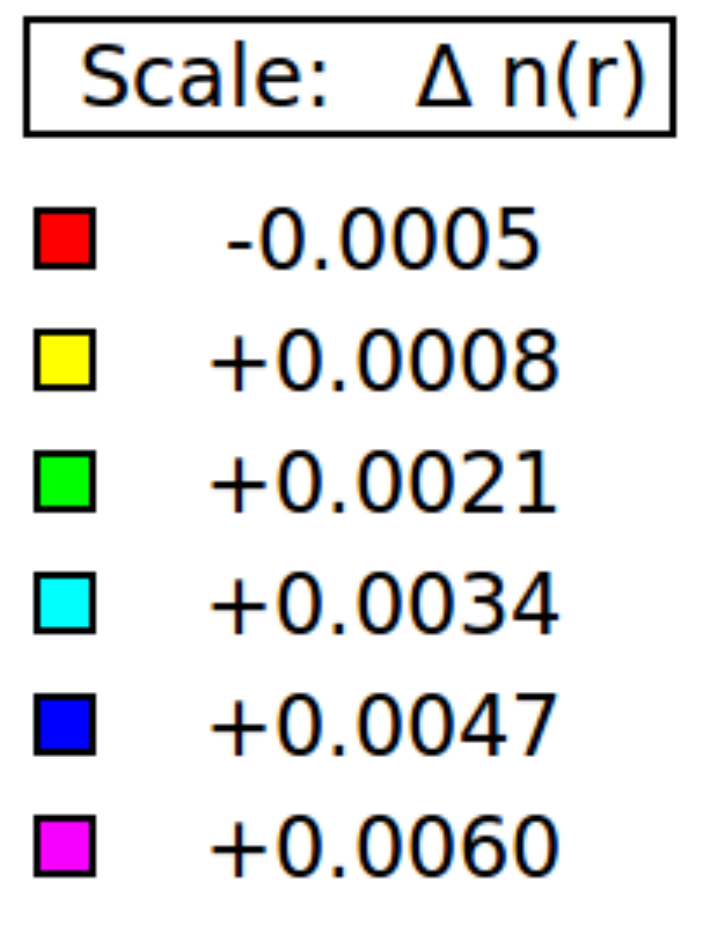}}
        \caption{The electronic density differential maps for graphene modified by the
molecule (a) and molecule modified by graphene (b).
The scale (right panel) is the same in both cases.}
        \label{Fig3}
\end{figure}

Adsorption of d-pentacene onto graphene is accompanied by charge transfer
from the molecule to the substrate. Figures \ref{F3a} and \ref{F3b} display the 
differential electronic density maps
within the plane of graphene and the d-pentacene central part, respectively. 
The charge density is the electronic density multiplied by -e.
These maps show difference between the actual electronic density of the system and 
that of the isolated substrate or molecule.  
The substrate is negatively charged, which agrees with previous studies
for a similar but smaller molecule (with the same dipole groups) placed between 
two graphene sheets \cite{x6}; in that case, the bottom layer was shown to play a role of the cathode.
Besides, the charge distribution within the substrate is quite delocalized. The change of the charge  
distribution induced in d-pentacene is positive in the central
part and negative around the dipole groups and adjacent carbons. This map supports the
scenario of separated paths for electrons and holes transport when the molecule is
between the graphene electrodes, as has been previously discussed for the smaller 
molecule \cite{x7}.  

The above results suggest that the physical behavior of d-pentacene adsorbed 
at graphene is affected by two effects: the geometry of the molecule, which changes 
upon adsorption, and the electrostatic interactions between the molecule and 
the substrate. Both effects are obviously related, since the geometrical changes 
are caused in turn by the electrostatic interactions. However, it is convenient to consider them separately 
to identify how they affect the electronic structure and 
the optical properties of the entire system. Therefore, we considered the three models described in the Materials and Methods section: Pi (isolated 
d-pentacene with its ``ideal'' plane geometry), PrG (d-pentacene adsorbed on graphene,
with its ``relaxed'' geometry), and Pr (isolated d-pentacene with the same geometry as for PrG).
Thus, comparison between Pi and Pr provides information about an effect of the geometry, 
and that between Pr and PrG accounts for the effect of weak $\pi$-bonds 
with the graphene substrate. 

\begin{figure}[h!]
\centerline{
     \includegraphics[scale=0.3,angle=0.0]{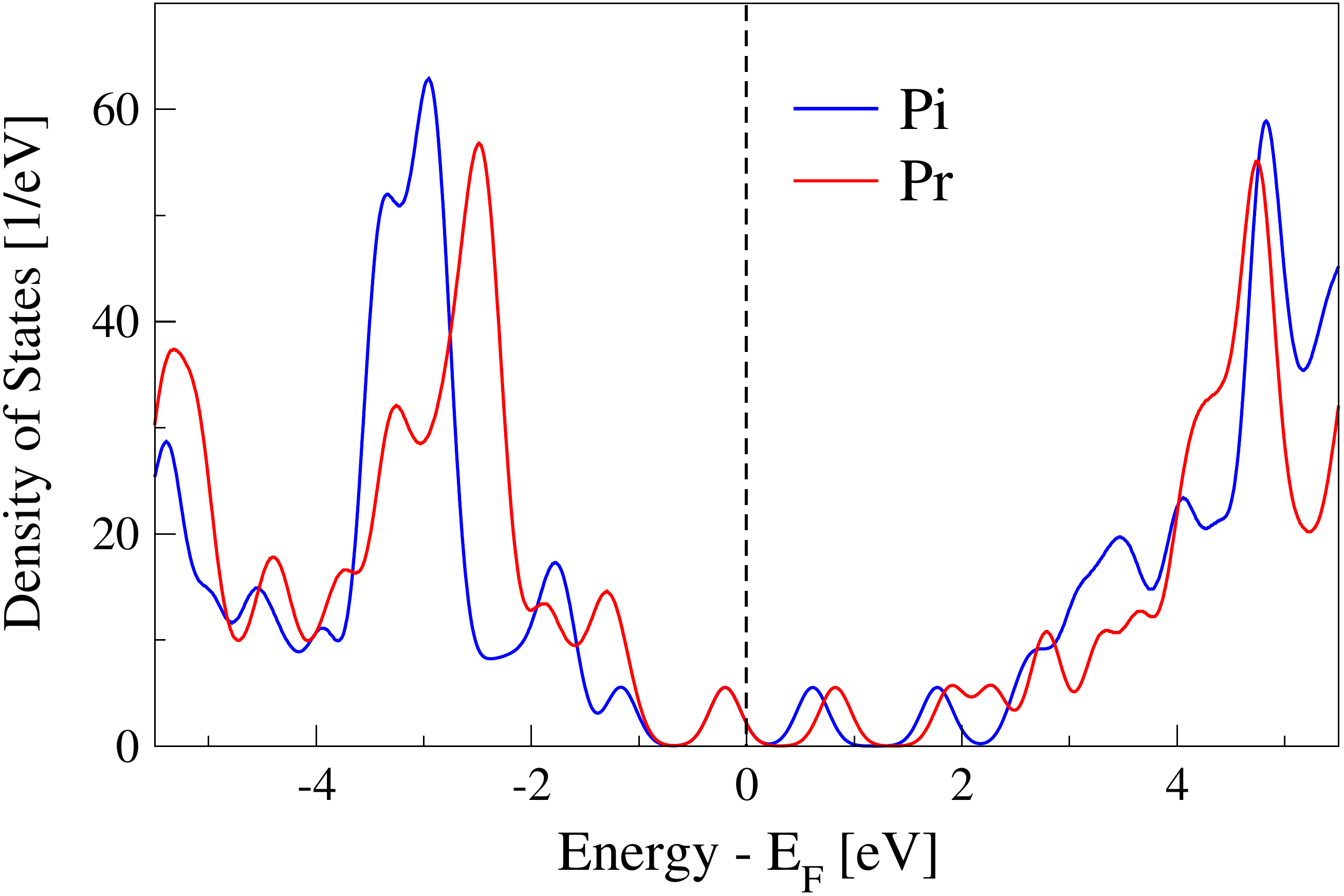}\label{Fig4-a}
     \includegraphics[scale=0.3,angle=0.0]{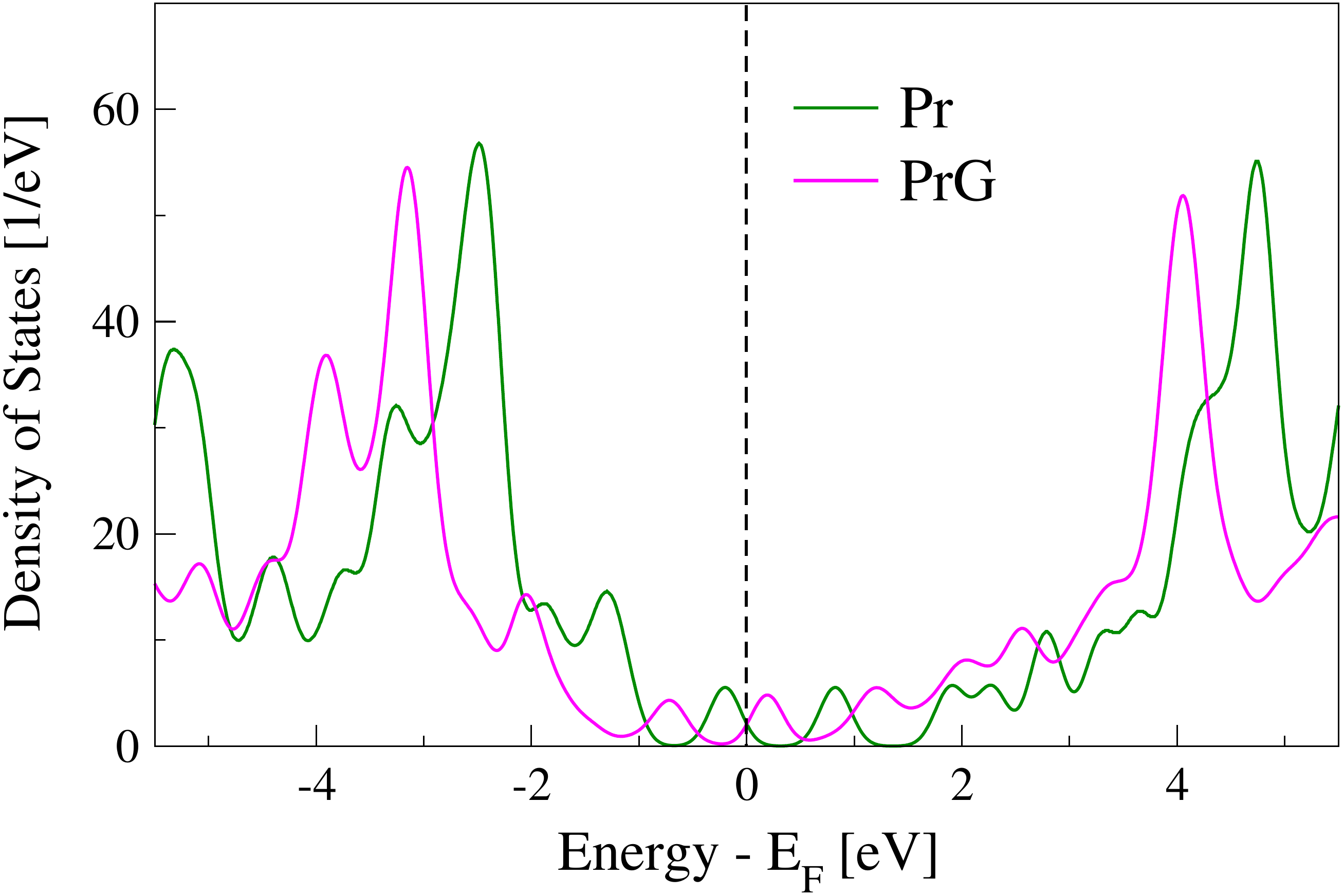}\label{Fig4-b}}

        \caption{Comparative densities of states: Pi and Pr configurations (a) and Pr and PrG ones (b).}
        \label{Fig4}
\end{figure}

Figure \ref{Fig4} shows the densities of states (DOS) for these three configurations; 
the origin of energies was taken as the corresponding Fermi energy for each case. 
Figure \ref{Fig4-a} plots the DOS for the Pi and Pr configurations. This 
figure gives an evidence that 
the geometry optimization of the molecule changes the distribution of the deep energy levels, 
but not those close to the HOMO--LUMO gap; in particular, the HOMO remains unchanged
(note that the red line overlaps with the blue one in Figure \ref{Fig4-a}). 
As for the LUMO, it is shifted upwards by about 0.2 eV, likely due to dipolar interactions 
within the molecule. In particular, the HOMO--LUMO gaps are $\varepsilon_{HO-LU}=0.81$ eV and 
$\varepsilon_{HO-LU}=1.02$ eV for Pi and Pr, respectively. Figure \ref{Fig4-a} also shows 
that the geometry optimization does not yield any charge transfer, since the Fermi energy 
remains unchanged relative to the HOMO. Figure \ref{Fig4-b}, on its part, compares 
the DOS for Pr and PrG cases. The electrostatic interactions with the substrate now shift 
upwards the Fermi energy, as one would expect given the donor character of the molecule 
that we have already mentioned. The remaining DOS is almost rigidly shifted downwards, what causes the system to become metallic. These plots demonstrate then that 
the changes in the molecular levels are mainly associated to the geometrical distortions. In essence, the effect of the substrate is the (roughly) rigid shift downwards of 
the entire DOS accompanying the charge transfer.
\begin{figure}[h!]
\centerline{
        \includegraphics[scale=0.3,angle=0.0]{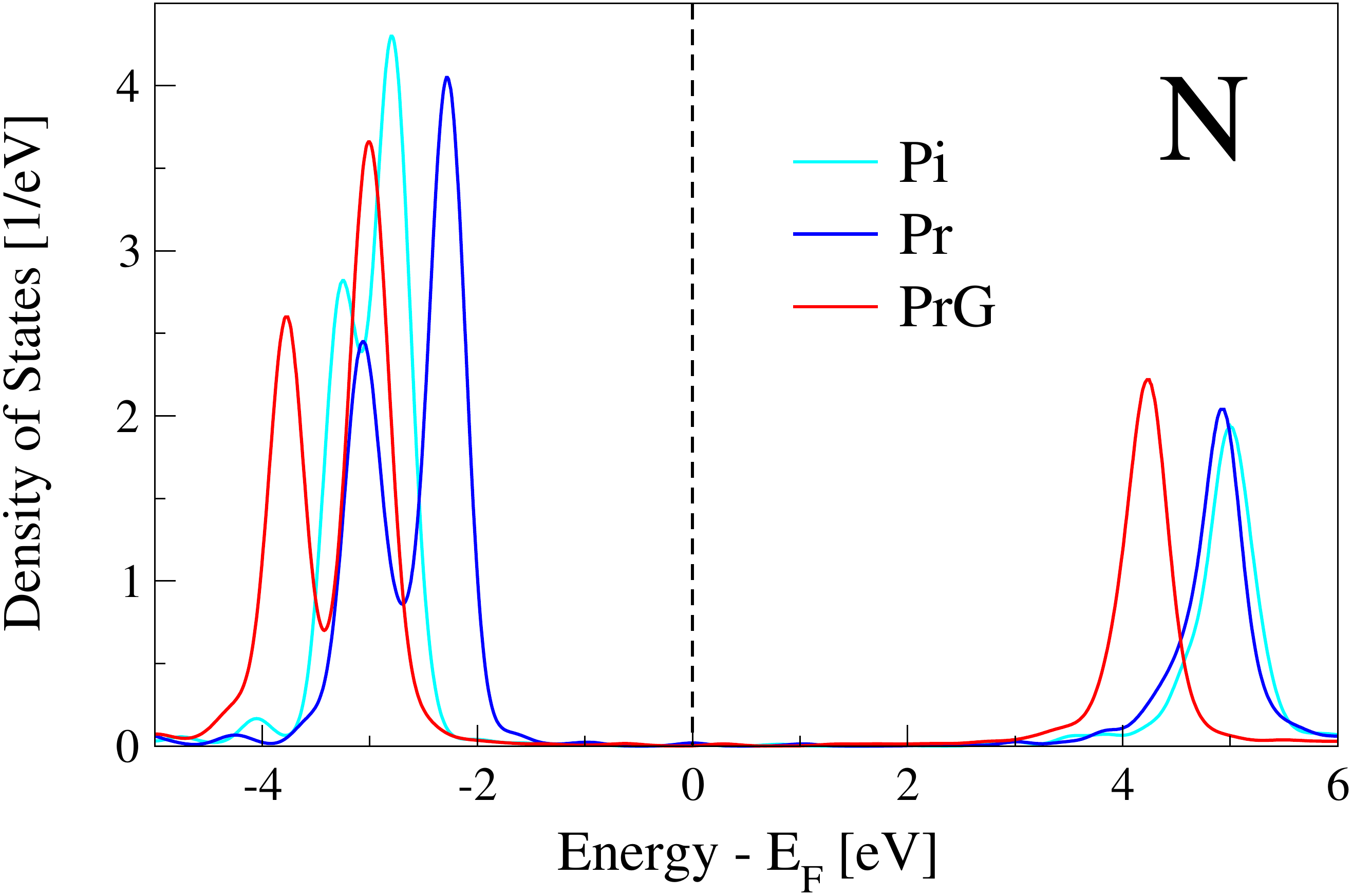}\label{Fig5-a}
        \includegraphics[scale=0.3,angle=0.0]{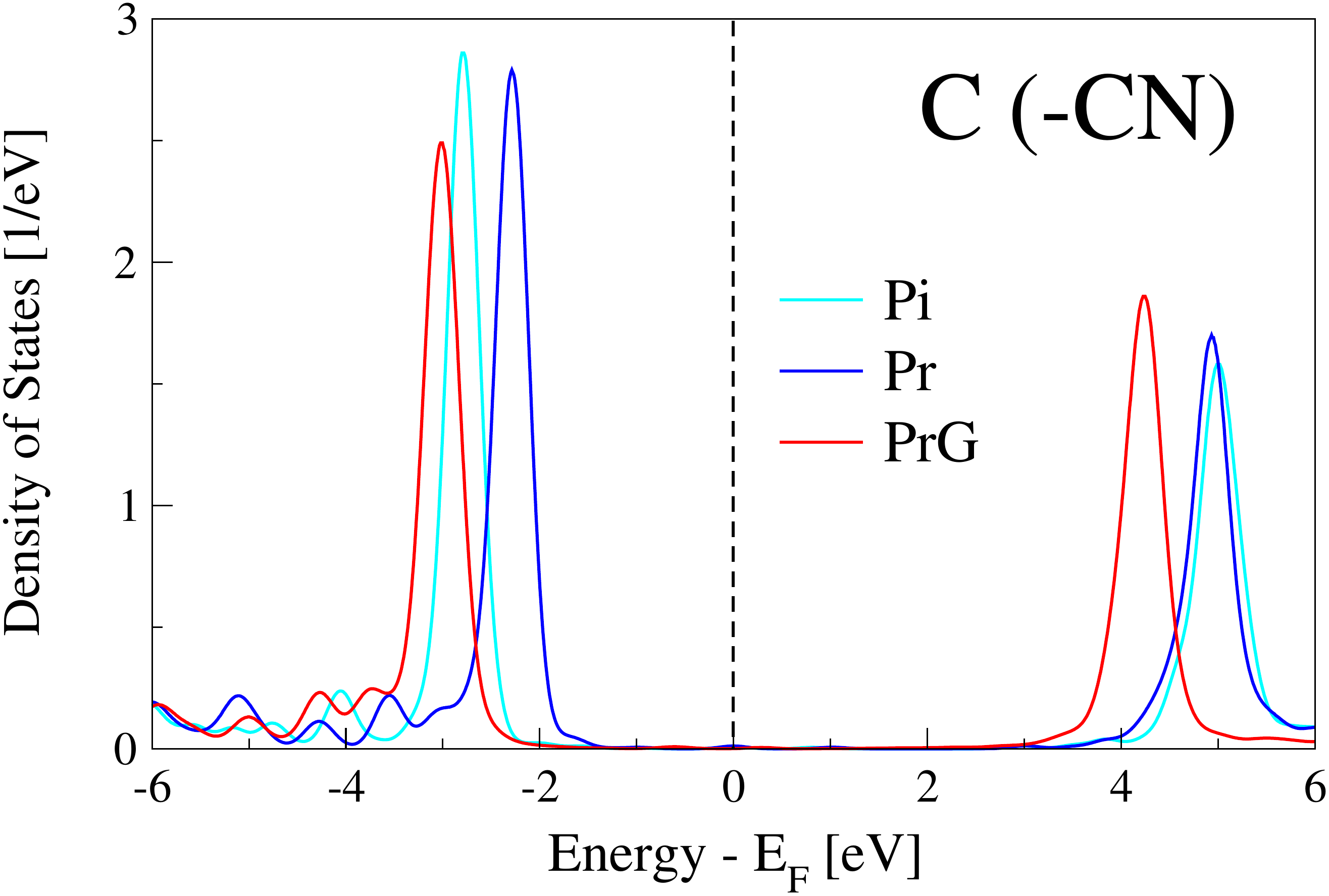}\label{Fig5-b}} 
\centerline{
        \includegraphics[scale=0.3,angle=0.0]{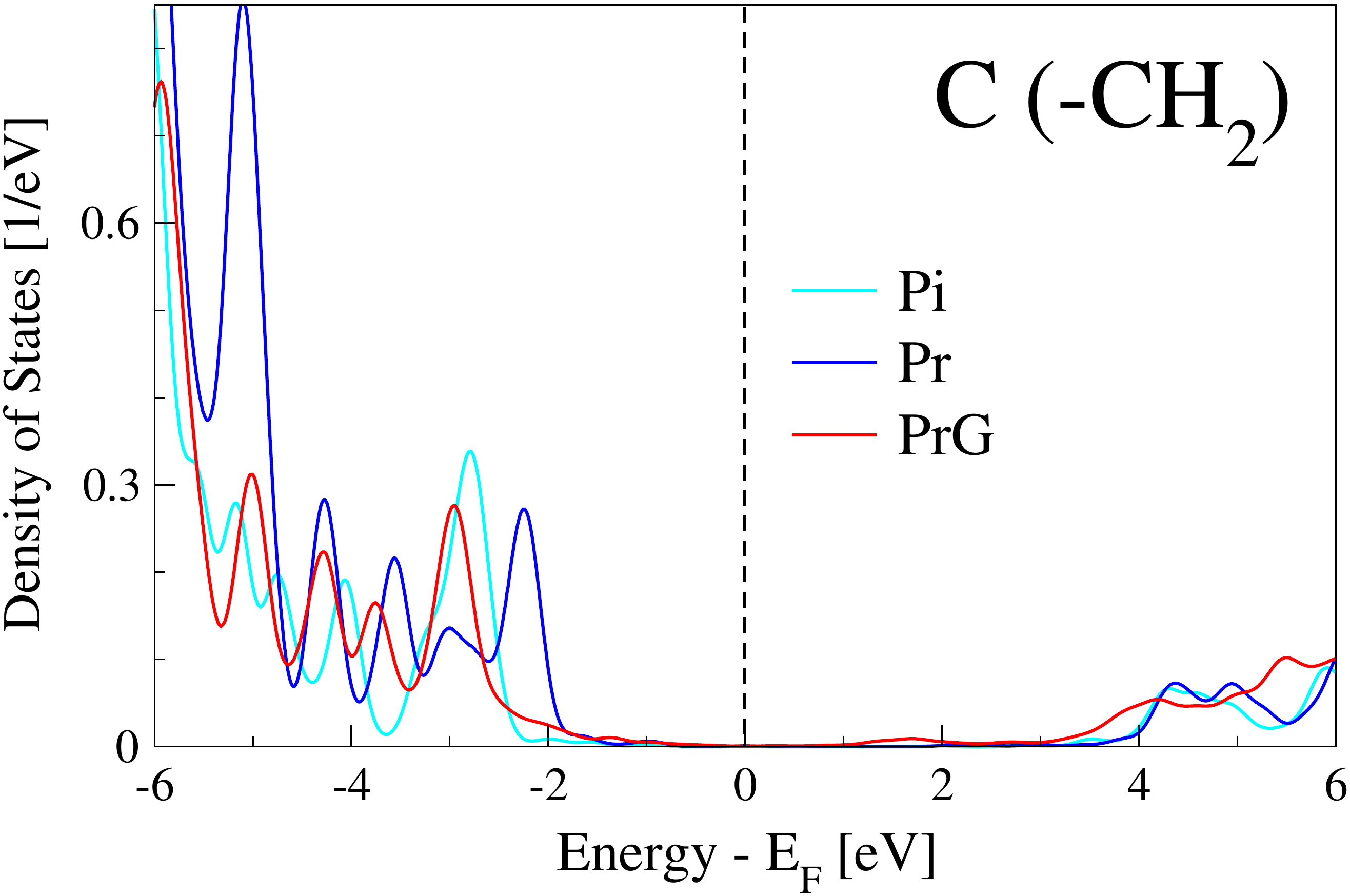}\label{Fig5-c}
        \includegraphics[scale=0.3,angle=0.0]{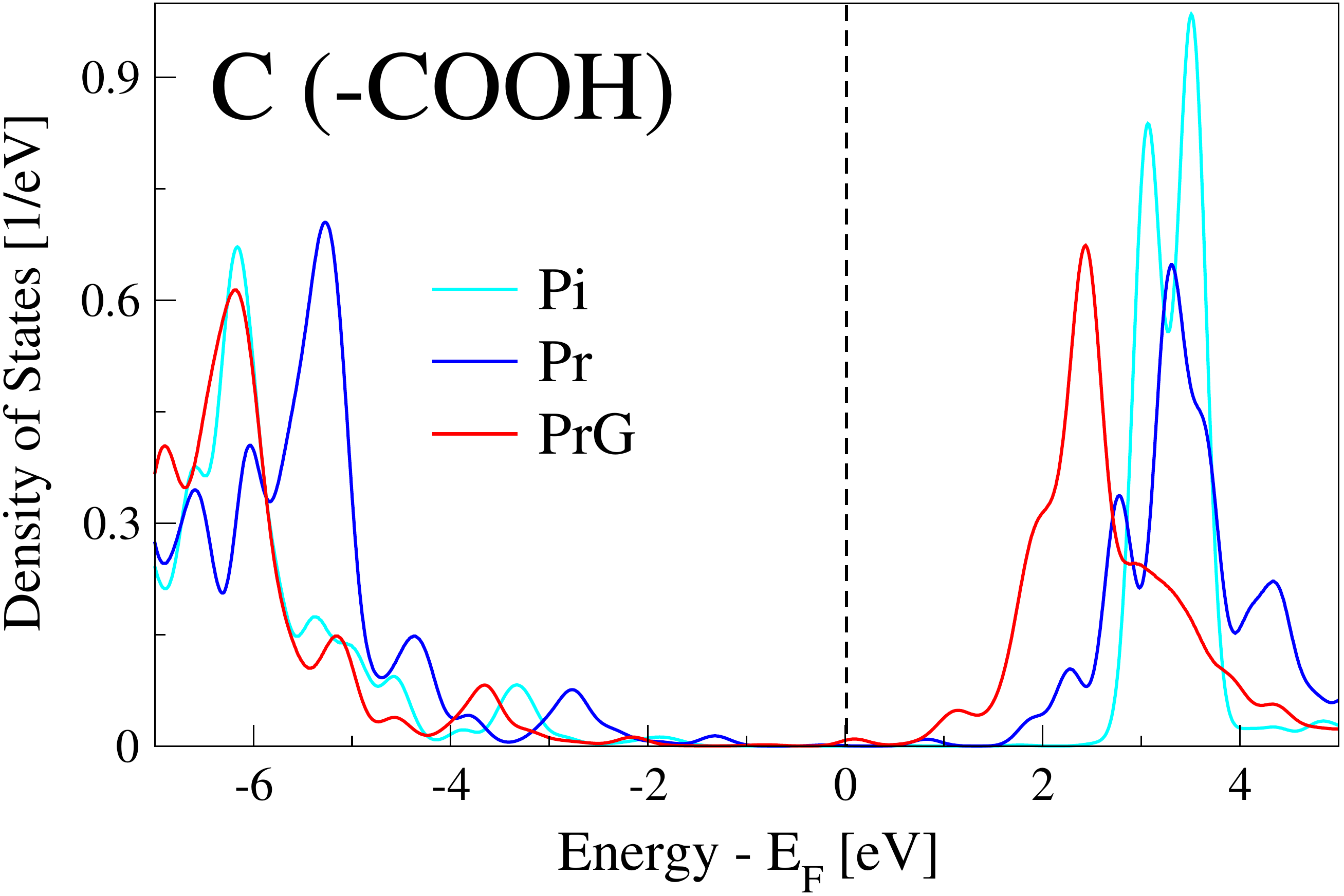}\label{Fig5-d}}

        \caption{Projected DOS at N (a) and C of --CN (b), --CH$_2$ (c), --COOH groups (d) for Pi, Pr and PrG.}
        \label{Fig5}
\end{figure}

Deeper insights into the electronic structure of the system are provided by the projection 
of the DOS onto selected atomic states. Figure \ref{Fig5} shows the projected DOS (pDOS) 
onto nitrogen (from the --CH$_2$CN group) and three distinct carbon (from the cyano, 
methylene and carboxyl groups) atoms for the Pi, Pr and PrG cases. Geometry relaxation does 
not affect much the projection of the unoccupied levels onto CH$_2$CN states 
(Figures \ref{Fig5-a} to \ref{Fig5-c}), whereas it causes occupied states to shift upwards by 
about 0.5 eV. On the contrary, the change in geometry affects greatly the projection onto 
the carbon atom from the carboxyl group. We observe in Figure \ref{Fig5-d} that the projections 
onto unoccupied levels become wider after the molecule gets distorted whilst the projection 
with lower energy shifts downwards. As for the projections onto occupied levels, they shift 
upwards almost rigidly by effect of the geometry optimization. On the other hand, 
the electrostatic interactions with the graphene layer cause essentially a rigid shift 
downwards of the projections in all cases.  

\begin{figure}[H]
\centerline{
        \includegraphics[scale=0.3,angle=0.0]{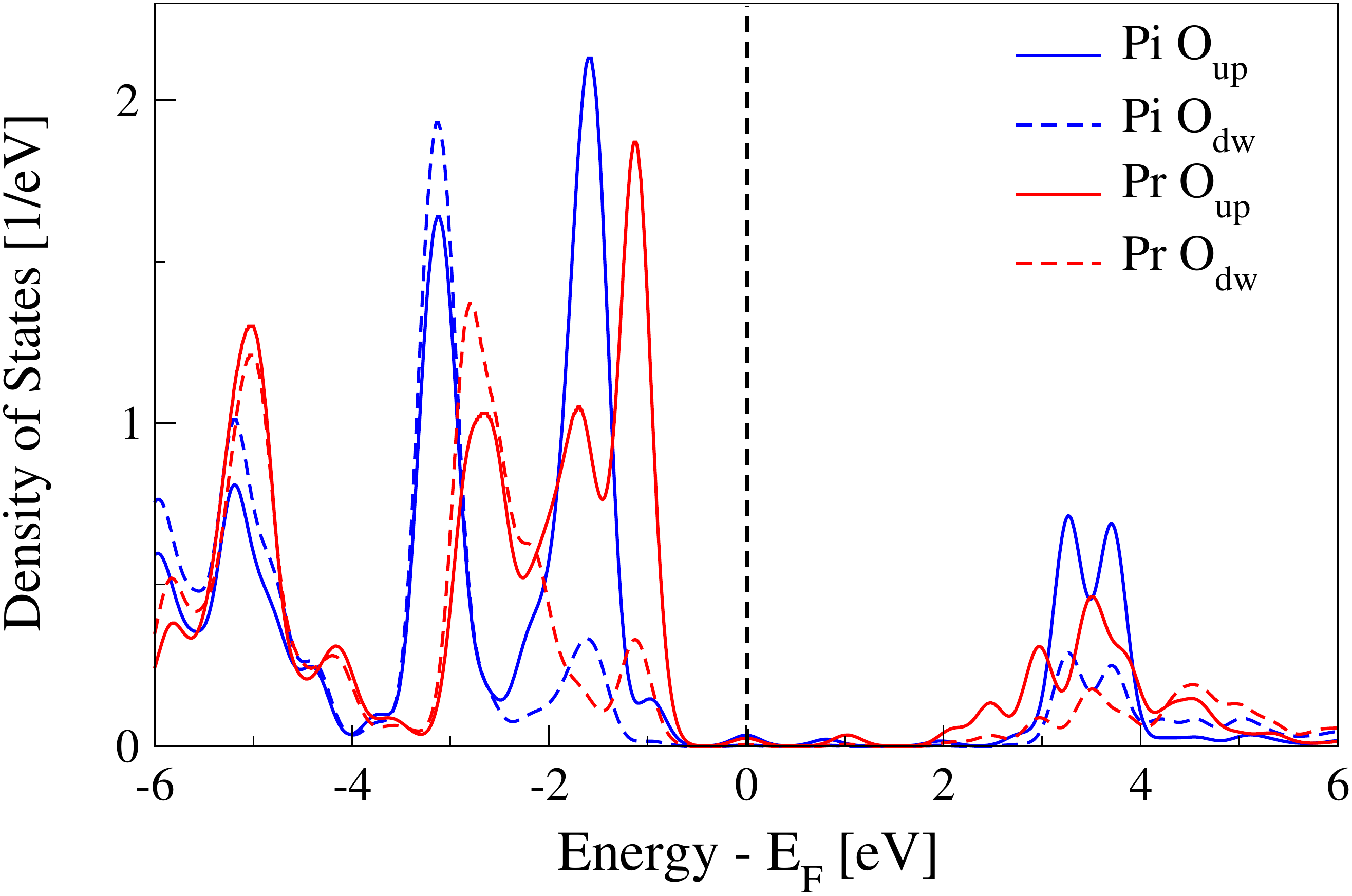}\label{Fig6-a}
        \includegraphics[scale=0.3,angle=0.0]{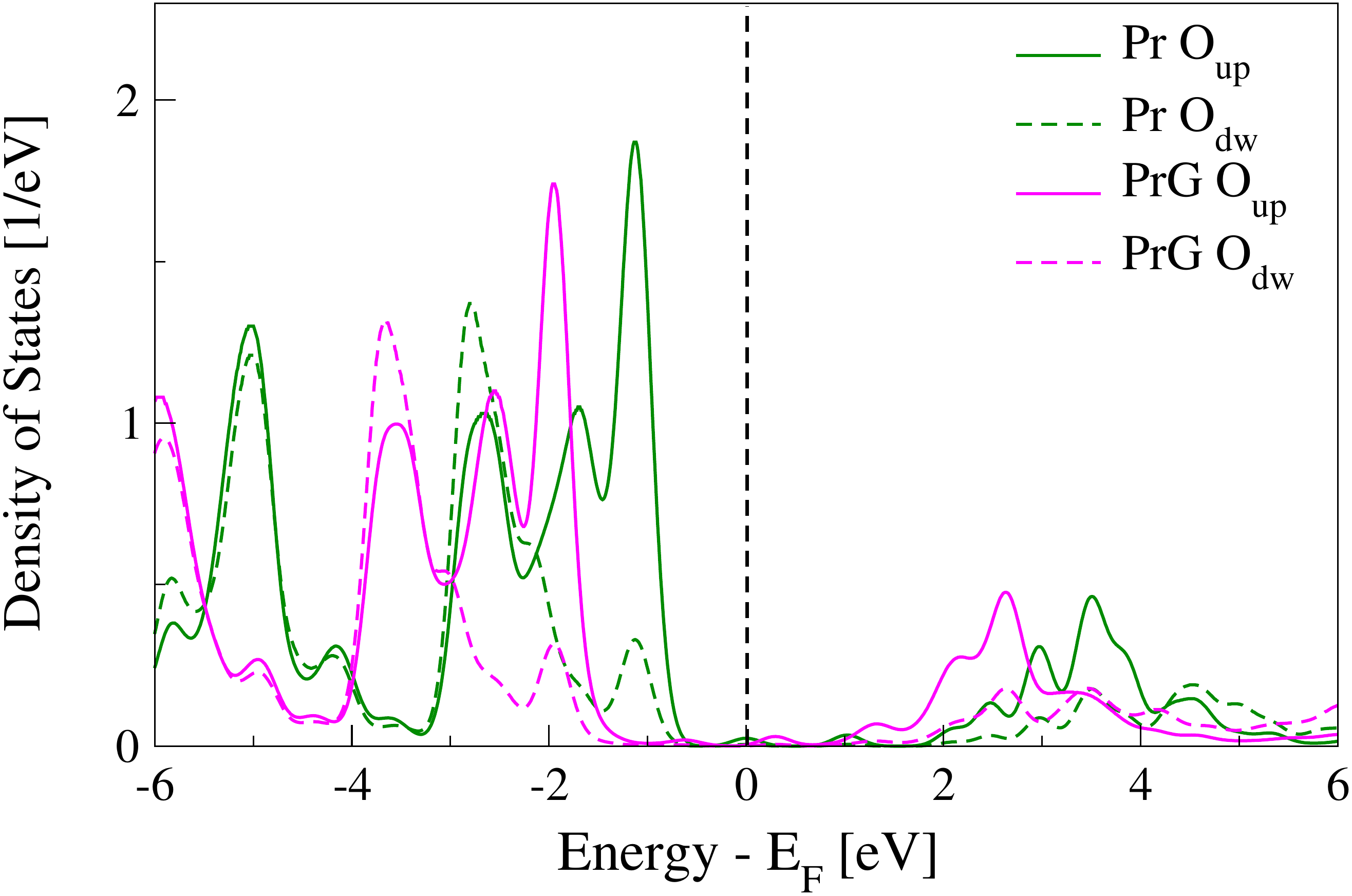}\label{Fig6-b}}
        \caption{PDOS onto oxygen states. 
Comparative between Pi and Pr (a) and Pr and PrG (b) configurations.}
        \label{Fig6}
\end{figure}

Figures \ref{Fig6-a} and \ref{Fig6-b} show the projection of the DOS 
onto non-equivalent oxygen atoms of 
the -COOH groups for the Pi, Pr and PrG configurations. One of these oxygens 
(labeled as O$_{up}$) is doubly bonded to the carbon atom, whereas the other (O$_{dw}$) 
is singly bounded to carbon and to a hydrogen atom as well. This is the reason why they 
show different chemical behavior. As for the previous figures, the main effect seems 
to come from the geometrical distortion of the molecule. Indeed, Figure \ref{Fig6-a} reveals 
a widening and shift downwards of the unoccupied levels, which is more pronounced for 
O$_{up}$, in Pr compared to Pi. On the contrary, and also as before, physisorption causes 
just a roughly shift downwards of the levels, with no apparent changes in their structure, 
as shown in Figure \ref{Fig6-b}. 
This is an indirect evidence that $\pi-$bonds are indeed established between the molecule 
and the graphene substrate. 

\subsection{Modified Chemical Structure}

\begin{figure}[h!]
	\centering
	\includegraphics[scale=0.6,angle=0.0]{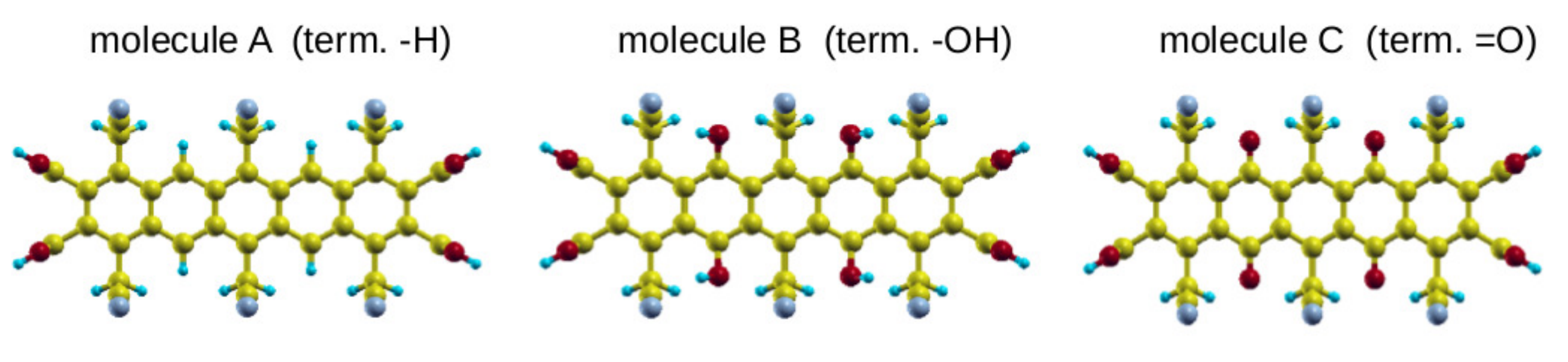}
	\caption{Atomic structure of the original d-pentacene (left) and its modifications: 
replacement of H with --OH (middle) and =O (right). The color code is the same as for Figure 1.}
	\label{Fig7}
\end{figure}

From the previous results, the question arises as to how to modify 
the electronic structure of 
the d-pentacene + graphene system to increase the optical efficiency through 
some ``levels engineering''. 
A quick response would be to change the dipole groups; one could 
then try different possibilities 
and check which one yields a suitable HOMO--LUMO gap. We have explored here a different approach: 
we have investigated an effect of changing the C--H single bonds within the d-pentacene molecule 
by either C--OH or C=O bonds. In the first case, even though one still have a single bond, 
the oxygen atom is likely to enrich the electrostatic interactions (within the molecule and between 
the molecule and the graphene sheet) because of its higher number of electrons. In the second case 
one has a double bond which ``fixes" the single-double bonds pattern 
of each aromatic ring, thus avoiding the Kekule-structures duality. In this case, the electrons 
can no longer be considered as delocalized within these rings, what incidentally decreases the
aromaticity. Figure \ref{Fig7} shows schematics of 
the original d-pentacene as well as its two modifications; in what follows, we will label them 
as molecules A, B and C, respectively. 

The proximity of the graphene layer to the modified molecules depend on their particular 
chemical conformation. For molecule B, the proximity of graphene causes additional perpendicular 
$\pi-$bonds to appear, and therefore a weakening of the double $\pi-$bonds in the planar 
mesogenic part of the molecule. For molecule C, we find weaker $\pi-$bonds in 
the mesogenic part, caused now by the double (instead of single) planar saturation bonds.

\begin{figure}[H]
\centerline{
      \includegraphics[scale=0.26,angle=0.0]{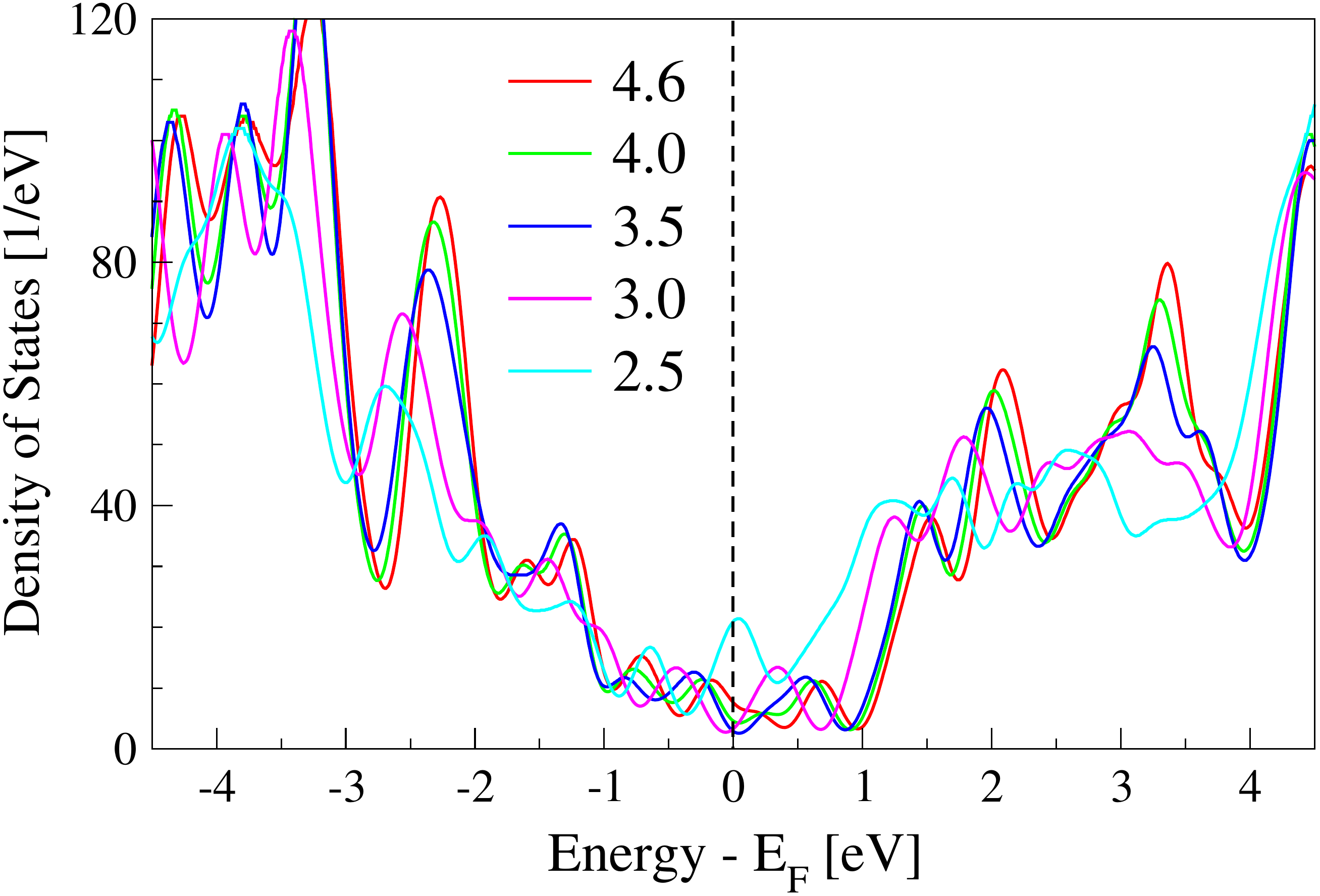}\label{Fig8a}
      \includegraphics[scale=0.26,angle=0.0]{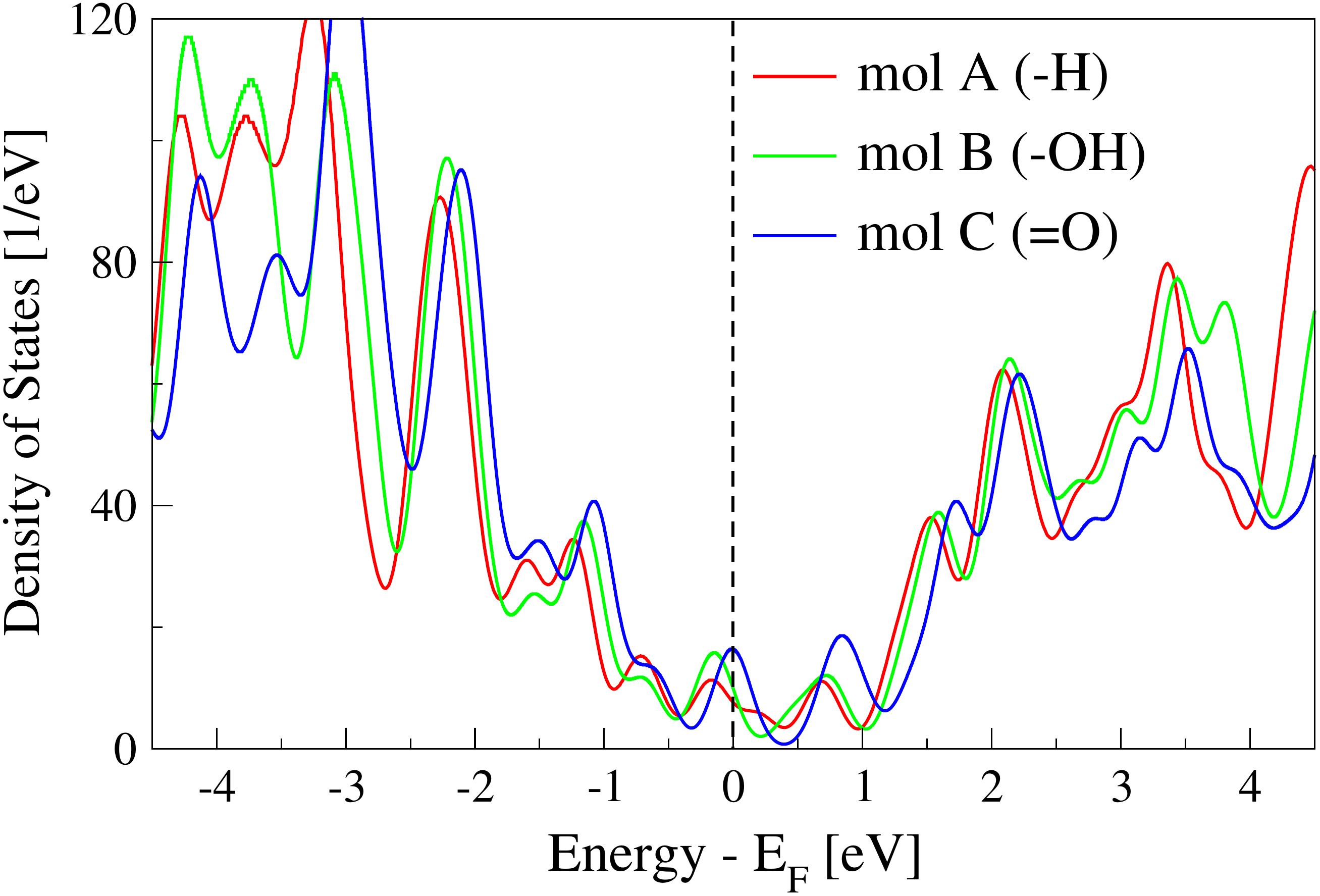}\label{Fig8b}}
        \caption{DOS of d-pentacene at graphene in various distances (a), and DOS of three
modifications of d-pentacene (see Figure \ref{Fig7}) (b).}
\end{figure}

Figures \ref{Fig8a} and \ref{Fig8b}
display the DOS of d-pentacene, in its ``ideal'' undistorted structure, at various distances
from graphene and a comparison of the electronic structures of molecules A, B, C located at
4.5 $\AA$ from graphene, respectively. 
Interestingly, a peak at the Fermi level appears when the original d-pentacene is very close 
to graphene (Figure \ref{Fig8a}) and when the molecule is modified with =O double bonds (Figure \ref{Fig8b}).    
Both effects, i.e., the formation of the perpendicular $\pi$-bonds and additional planar $\pi$-bonds,
cause the aforementioned weakening of the bonds in the mesogenic part. 
Consequently, as we expected, the chemical 
modifications of the terminal parts can be used for engineering of the optical spectra. 

\begin{figure}[h!]
\centerline{
      	\includegraphics[scale=0.75,angle=0.0]{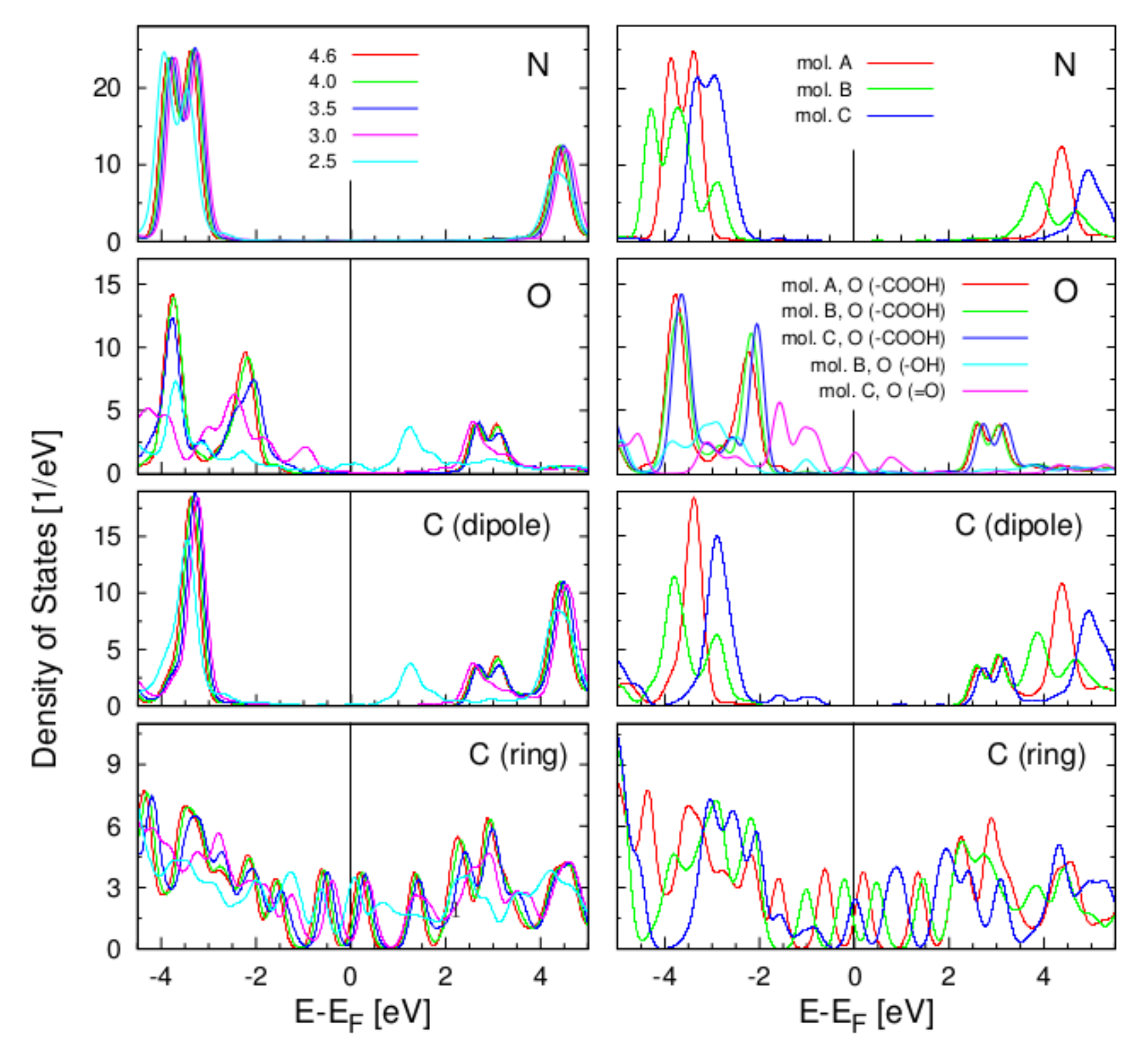}}
	\caption{The density of states projected at N, O, and C of the dipoles and 
the central rings for the molecule A in various distances from graphene (left panels) 
and the same projections for the molecules A, B, C at graphene in the distance 4.6 
$\AA$ (right panels). The contributions of the O-states from the terminal moieties --OH 
and =O of the molecules B and C, respectively, are also displayed.}
	\label{Fig9}
\end{figure}

Figure \ref{Fig9} shows the DOS projected onto selected parts of the A, B and C molecules, as well
as the molecule A in various distances from graphene.
According to our intuitive guess, the non-zero DOS values at $\varepsilon_F$ arise from 
the C atoms of the central rings. A small contribution to the DOS at $\varepsilon_F$ 
from the --COOH group, as well as contribution of
the additional oxygens for molecule C, appears 
when the molecule and graphene are about 2.5 $\AA$ away. 
For the most distant molecular part, namely the cyano group, 
the charge redistribution due to the perpendicular 
$\pi$-bonds with graphene does not affect the pDOS. This is in contrast to the terminal chemical 
modifications, which shift the peaks and change
their width for both the occupied and unoccupied N-states.

\subsection{Optical Properties}

\begin{figure}[h!]
        \begin{center}
                \includegraphics[scale=0.4,angle=0.0]{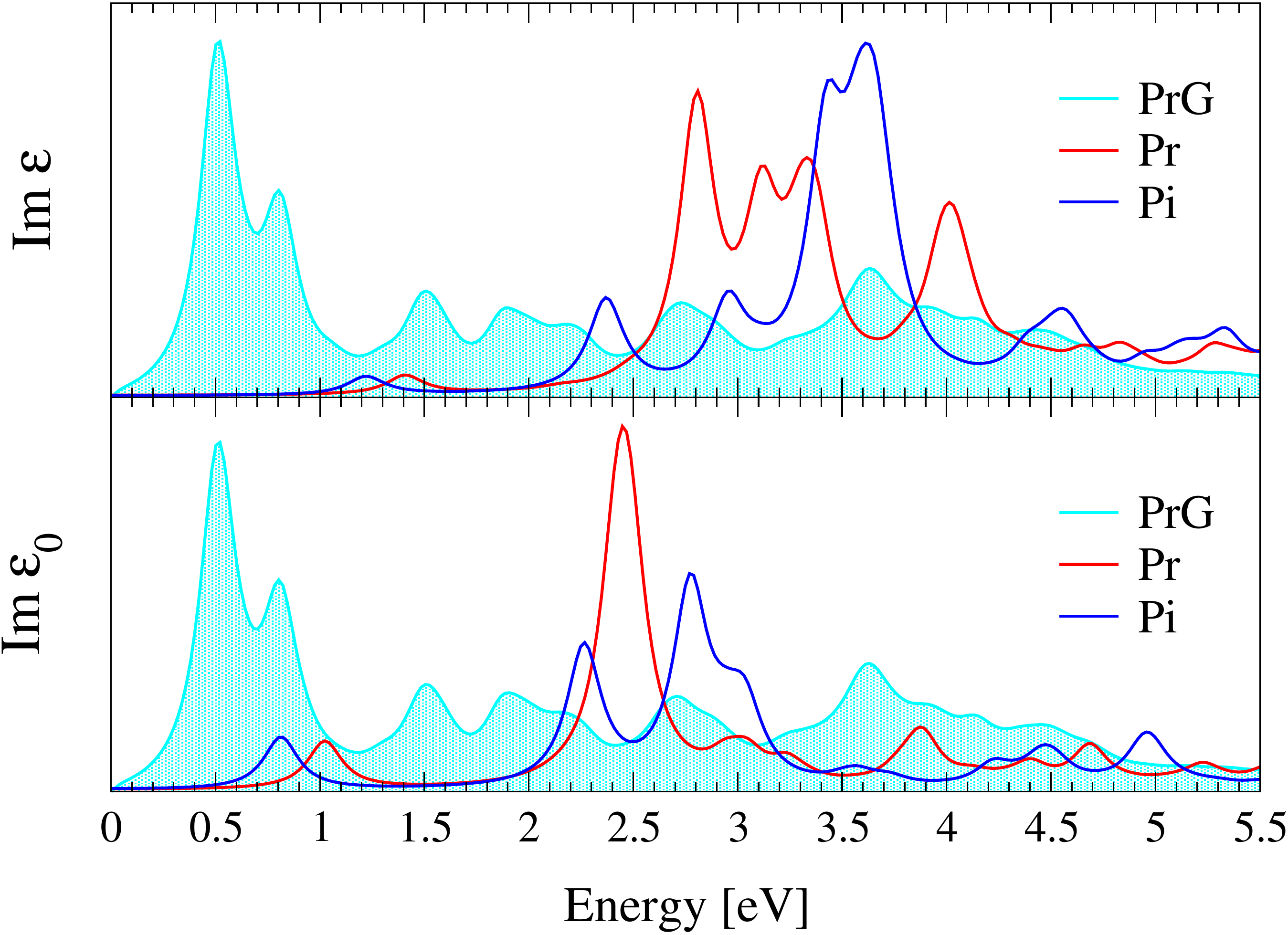}
                \caption{The interacting ($\varepsilon$) and noninteracting ($\varepsilon_0$)
                dielectric functions for Pi, Pr and PrG.}
                \label{Fig10}
                \end{center}
\end{figure}

The optical properties of isolated d-pentacene in its ``ideal" and ``relaxed" geometries 
(that is, Pi and Pr, respectively), as well as those of the molecule adsorbed at 
graphene (PrG), were studied within the linear response theory under the RPA. 
Figure \ref{Fig10} (upper panel) shows the imaginary part of the RPA or ``interacting'' 
dielectric function $\varepsilon(\omega)$. This imaginary part is proportional to 
the optical absorption coefficient. For comparison purposes, we computed 
the non-interacting, that is, corresponding to a system of independent particles 
dielectric function $\varepsilon_0(\omega)$ as well; this is shown in Figure \ref{Fig10} (bottom panel).

The first (lowest energy) peaks of the noninteracting dielectric function of isolated molecules 
correlate with the corresponding HOMO--LUMO gaps, 0.8 eV for Pi and 1.0 eV for Pr. 
The pDOS in this energetic region suggests that the optical transitions occur between 
the pentacene-chain localized states. These peaks have very low intensities, 
what is probably related to the dipole-symmetry-based transition rules. 
The group of intense peaks in the noninteracting dielectric function, roughly between 
2.0 and 4.0 eV, could correspond to excitations from valence band states localized at 
dipole-groups to conduction band states correlated to d-pentacene rings, as well as 
from valence band states centered at the d-pentacene core to conduction band states centered 
at dipoles. In the PDOS, most of Pi high-density states are further from the Fermi energy 
than the corresponding states for Pr, although some low-density states are closer 
to $\varepsilon_F$. These features correlate with the noninteracting dielectric functions 
of the appropriate cases. 

Compared to its noninteracting counterpart, and in very general sense, the peaks of 
the interacting dielectric function are generally wider than those of $\varepsilon_0$. 
For the Pi configuration, the peak exhibited by $\varepsilon_0$ at 2.3 eV remains 
unchanged in $\epsilon$; however, peaks at higher energy shift to the blue light. 
More interesting is the situation for Pr. In this case, it is the peak at around 4.0 eV 
which remains roughly unchanged, whereas the intense peak of $\varepsilon_0$ at 2.5 eV 
is splitted into three peaks within the 2.8--3.4 eV range. 
We remark two additional characteristics. The first is that the peaks for the HOMO--LUMO 
transitions, which are little intense, but apparent in the noninteracting case, 
do not appear in the interacting spectrum, which indicates that the direct transitions 
mediated by the screened potential are forbidden (we recall that we are accepting the RPA). 
The second characteristic is that, contrarily to the non-interacting case, 
molecular distortions cause a shift to the blue light, by about 0.5 eV, in the interacting case.

\begin{figure}[h!]
        \begin{center}
        \includegraphics[scale=0.4,angle=0.0]{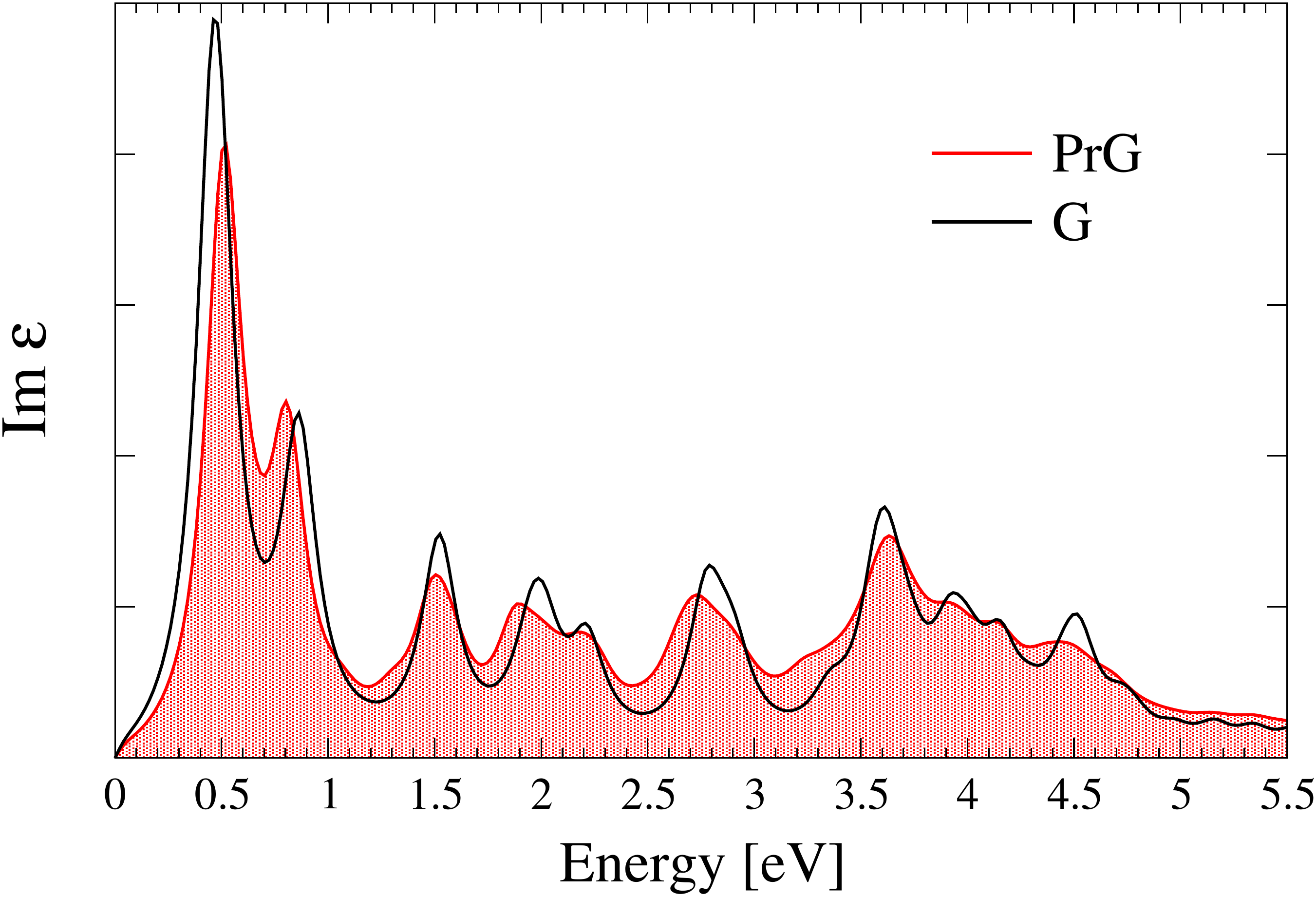}
        \caption{Comparison between the interacting dielectric function of PrG 
and that of graphene calculated in the same cell as PrG.}
        \label{Fig11}
        \end{center}
\end{figure}

The spectrum of d-pentacene at graphene (PrG) does not vary much between the noninteracting 
and interacting cases, as shown in Figure \ref{Fig10}, and it does not show prominent features 
which could be assigned to the electrostatic interactions between the molecule and 
the substrate either.  Moreover, the graphene-based transitions are so rich that 
it is difficult to find the molecular and the molecule-to-graphene transitions. 
This is demonstrated in Figure \ref{Fig11}, where the imaginary part of the interacting 
dielectric function of the PrG configuration is compared to that of the graphene layer 
(without any adsorbed molecule). This plot shows that the spectrum of PrG practically mimics 
that of pure graphene. This is an effect of the large number of atoms in graphene substrate, 
and therefore the electronic states, which are optically active. However, Figure \ref{Fig11} 
shows a slight supression of the lowest transitions within the substrate. 
Thus, the optical transitions between d-pentacene and graphene seem to cause a slight 
shift to the right for the low-energy part of the PrG spectrum, as well as 
the intensification of this spectrum in the region below the HOMO--LUMO gap of the molecule.

In any case, the analysis of the optical response is consistent with our previous discussion in the sense that the most remarkable physical change in the system arises from the molecular distortions, and not from interactions with the graphene substrate. Of course, the analysis of optical properties in the system is far to be conclusive, since the energy levels have been calculated within ``standard'' DFT, ignoring many-body corrections. In particular, the possibility of excitonic resonances is not taken into account in RPA. However, this study could open a research field not only about the role of intermolecular interactions in this type of system, but also about possible ``levels engineering'' strategies.

\section{Conclusions}

The spectroscopic properties of a photovoltaic molecule, namely pentacene decorated with 
four COOH and six CH$_2$CN dipole groups and adsorbed at a graphene substrate have been 
studied by {\it ab initio} methods: DFT and RPA for the dielectric function.

Our results show that chemisorption does not occur; instead, physisorption accompanied by 
a ``cone"-like distortion of the molecule takes place. A charge transfer from d-pentacene 
to graphene is correlated with charge redistribution within the molecule, where the central 
part yields electrons to the dipole terminal groups. In general terms, the effect of 
the molecular distortion shifts the electronic spectrum upwards, while the weak $\pi$-bonding 
with the substrate manifests itself in the opposite way.

The chemical modification of the saturation from --H terminal to the electron-rich --OH and =O 
terminals changes the single-double bond pattern in the central part of d-pentacene, 
leading to smaller HOMO--LUMO gaps. The same effect is observed for the $\pi$-bonds between 
the molecule and graphene at small distances. This observation can be used in the ``engineering'' 
of the optical spectra. It is remarkable that these molecules have their lowest 
appreciable transitions at relatively high energy, what makes them not optimal for 
efficient photovoltaics. 

Trends in the features of the dielectric function correspond to those drawn from the projected 
density of states. In particular, one sees that important changes appear in the optical 
absorption spectrum after the molecule gets distorted. However, adsorption of the molecule 
does not result in significant changes in the optical absorption spectrum with respect to that 
for single graphene. Further discussion of the optical behaviour would require more 
sophisticated theoretical schemes which would include particle-hole interactions. 

\section*{Acknowledgments}
Calculations have been performed in the Cyfronet Computer Centre using the Prometheus computer,
which is a part of the PL-Grid Infrastructure.
This work has been supported by The National Science Centre of Poland
(the Projects: 2013/11/B/ST3/04041 and DEC-2012/07/B/ST3/03412) 
and by the Junta of Extremadura through Grant GR15105.

\section*{Conflict of Interest}
The authors declare that there is no conflict of interest regarding the publication of this paper.

\end{document}